\documentclass[aps,prl,twocolumn,superscriptaddress,floatfix]{revtex4-1}

\usepackage[latin9]{inputenc}
\usepackage{graphicx}
\usepackage{dcolumn}
\usepackage{bm}
\usepackage{amssymb}
\usepackage{microtype}
\usepackage{xfrac}
\usepackage{makecell}
\usepackage{array}
\usepackage[version=4]{mhchem}
\usepackage{gensymb}
\usepackage{multirow}
\usepackage[colorlinks=true]{hyperref}
\begin{document}
\title{Intertwined Charge and Spin Density Waves in a Topological Kagome Material}

\author{Y.~Chen}
\affiliation{Institute for Quantum Matter and Department of Physics and Astronomy, Johns Hopkins University, Baltimore, MD 21218, USA}
\affiliation{Department of Physics, University of California, Berkeley, CA 94720, USA}
\affiliation{Material Sciences Division, Lawrence Berkeley National Lab, Berkeley, California 94720, USA}

\author{J.~Gaudet}
\affiliation{Institute for Quantum Matter and Department of Physics and Astronomy, Johns Hopkins University, Baltimore, MD 21218, USA}
\affiliation{NIST Center for Neutron Research, National Institute of Standards and Technology, Gaithersburg, Maryland 20899, USA}
\affiliation{Department of Materials Science and Eng., University of Maryland, College Park, MD 20742-2115}

\author{G.~G.~Marcus}
\affiliation{Institute for Quantum Matter and Department of Physics and Astronomy, Johns Hopkins University, Baltimore, MD 21218, USA}

\author{T.~Nomoto}
\affiliation{Research Center for Advanced Science and Technology, University of Tokyo, 4-6-1 Komaba Meguro-ku, Tokyo 153-8904, Japan }

\author{T.~Chen}
\affiliation{Institute for Solid State Physics (ISSP), University of Tokyo, Kashiwa, Chiba 277-8581, Japan}

\author{T.~Tomita}
\affiliation{Institute for Solid State Physics (ISSP), University of Tokyo, Kashiwa, Chiba 277-8581, Japan}

\author{M.~Ikhlas}
\affiliation{Institute for Solid State Physics (ISSP), University of Tokyo, Kashiwa, Chiba 277-8581, Japan}

\author{H.~S.~Suzuki}
\affiliation{Institute for Solid State Physics (ISSP), University of Tokyo, Kashiwa, Chiba 277-8581, Japan}

\author{Y.~Zhao}
\affiliation{NIST Center for Neutron Research, National Institute of Standards and Technology, Gaithersburg, Maryland 20899, USA}
\affiliation{Department of Materials Science and Eng., University of Maryland, College Park, MD 20742-2115}

\author{W.~C.~Chen}
\affiliation{NIST Center for Neutron Research, National Institute of Standards and Technology, Gaithersburg, Maryland 20899, USA}

\author{J.~Strempfer}
\affiliation{Advanced Photon Source, Argonne National Laboratory, Illinois 60439, USA}

\author{R.~Arita}
\affiliation{Research Center for Advanced Science and Technology, University of Tokyo, 4-6-1 Komaba Meguro-ku, Tokyo 153-8904, Japan }
\affiliation{RIKEN Center for Emergent Matter Science, 2-1 Hirosawa Wako Saitama 351-0198, Japan}

\author{S.~Nakatsuji}
\affiliation{Institute for Quantum Matter and Department of Physics and Astronomy, Johns Hopkins University, Baltimore, MD 21218, USA}
\affiliation{Institute for Solid State Physics (ISSP), University of Tokyo, Kashiwa, Chiba 277-8581, Japan}
\affiliation{Department of Physics, University of Tokyo, Bunkyo-ku, Tokyo 113-0033, Japan}
\affiliation{Trans-scale Quantum Science Institute, University of Tokyo, Bunkyo-ku, Tokyo 113-8654, Japan}
\affiliation{Canadian Institute for Advanced Research, Toronto, M5G 1Z7, ON, Canada}

\author{C.~Broholm}
\affiliation{Institute for Quantum Matter and Department of Physics and Astronomy, Johns Hopkins University, Baltimore, MD 21218, USA}
\affiliation{NIST Center for Neutron Research, National Institute of Standards and Technology, Gaithersburg, Maryland 20899, USA}
\affiliation{Department of Materials Science and Engineering, Johns Hopkins University, Baltimore, MD 21218, USA}

\begin{abstract} 
Using neutrons and x-rays we show the topological kagome antiferromagnet Mn$_3$Sn for $T<285$~K forms a homogeneous spin and charge ordered state comprising a longitudinally polarized spin density wave (SDW) with wavevector $\textbf{k}_{\beta}=k_\beta {\bf \hat{c}}$, a helical modulated version of the room temperature anti-chiral magnetic order with $\textbf{k}_{\chi}=k_\chi{\bf \hat{c}}$, and charge density waves with wave vectors $2\textbf{k}_\beta, 2\textbf{k}_\chi$, and $\textbf{k}_\beta+\textbf{k}_\chi$. Though $\textbf{k}_{\chi}$ and $\textbf{k}_\beta$ coincide for $200~{\rm K}<T<230$~K, they exhibit distinct continuous $T-$dependencies before locking to commensurate values of $\textbf{k}_{\beta} = \frac{1}{12}\textbf{c}^{*}$ and $\textbf{k}_{\chi} = \frac{5}{48}\textbf{c}^{*}$ at low$-T$. Density functional theory indicates this complex modulated state may be associated with the nesting of Fermi surfaces from correlated flat kagome bands, which host Weyl nodes that are annihilated as it forms.
\end{abstract}

\date{\today}
\maketitle
The electronic band structure of kagome metals feature van Hove singularities and flat electronic bands wherein electronic interactions can produce strongly correlated and topologically non-trivial states of matter. In the spin sector, kagome materials have been shown to form both ferro\cite{liu2018giant} and antiferromagnetic\cite{Nakatsuji2015} Weyl semimetals\cite{Armitage2018}, while in the charge sector superconductivity and charge density waves have been discovered in the $\rm AV_3Sb_5$ (A = K, Rb, Cs) family of materials\cite{WOS:000486663000001,Jiang2021,PhysRevMaterials.5.034801,ortiz2020cs,PhysRevX.11.031050}. Further strong correlation is expected to lead to a nontrivial order that encompasses multiple electronic degrees of freedom, similar to those seen in various oxide materials~\cite{Moreo1999}. However, no such intertwined order has been reported for kagome metals to date. Here, we report that the low-temperature phase of the antiferromagnetic Weyl semimetals Mn$_3$Sn~\cite{Kuroda2017,chen2021anomalous} that was previously thought to be a heterogeneous mixture of two helical magnets\cite{CABLE1993,Song2020}, is actually a homogeneous incommensurate spin and charge modulated phase with two fundamental wave vectors that evolve continuously from incommensurate to commensurate upon cooling. 


For temperatures between 285 K and 420 K~\cite{tomiyoshi1982,Tomiyoshi1982v2,OHMORI1987,Brown1990}, Mn$_3$Sn forms an anti-chiral $\bf k=0$ magnetic structure, which like a ferromagnet, breaks time reversal and rotation symmetries. Though the uniform magnetization of Mn$_3$Sn is three order of magnitude smaller than in conventional ferromagnets, the material exhibits large anomalous Hall and Nernst effects with considerable application potential~\cite{Nakatsuji2015,Ikhlas2017,Kimata2019}. However, our primary focus here lies in the long-wavelength modulated magnetic phase observed for temperatures below 285 K. Using polarized neutrons, we show two wave vectors $\textbf{k}_{\chi}$ and $\textbf{k}_{\beta}$ are associated with kagome-plane polarized helical order and the out-of-plane polarized spin density waves (SDW), respectively. The observation of scattering from higher harmonics unambiguously proves these distinct components form a homogeneous spin and charge ordered state. Using synchrotron x-ray diffraction, we furthermore discovered that these spin structures are accompanied by charge density waves (CDW)s. We associate this complex order with nesting instabilities involving topologically non-trivial flat electronic bands, associated with the underlying kagome structure~\cite{Li2018,ghimire2020}. Using inelastic magnetic neutron scattering, we show that the triplet of spin wave excitations expected for an isotropic triangular-based AFM is split into three distinct modes in the low-temperature modulated phase.

\begin{figure}[t]
    \includegraphics[width=\columnwidth]{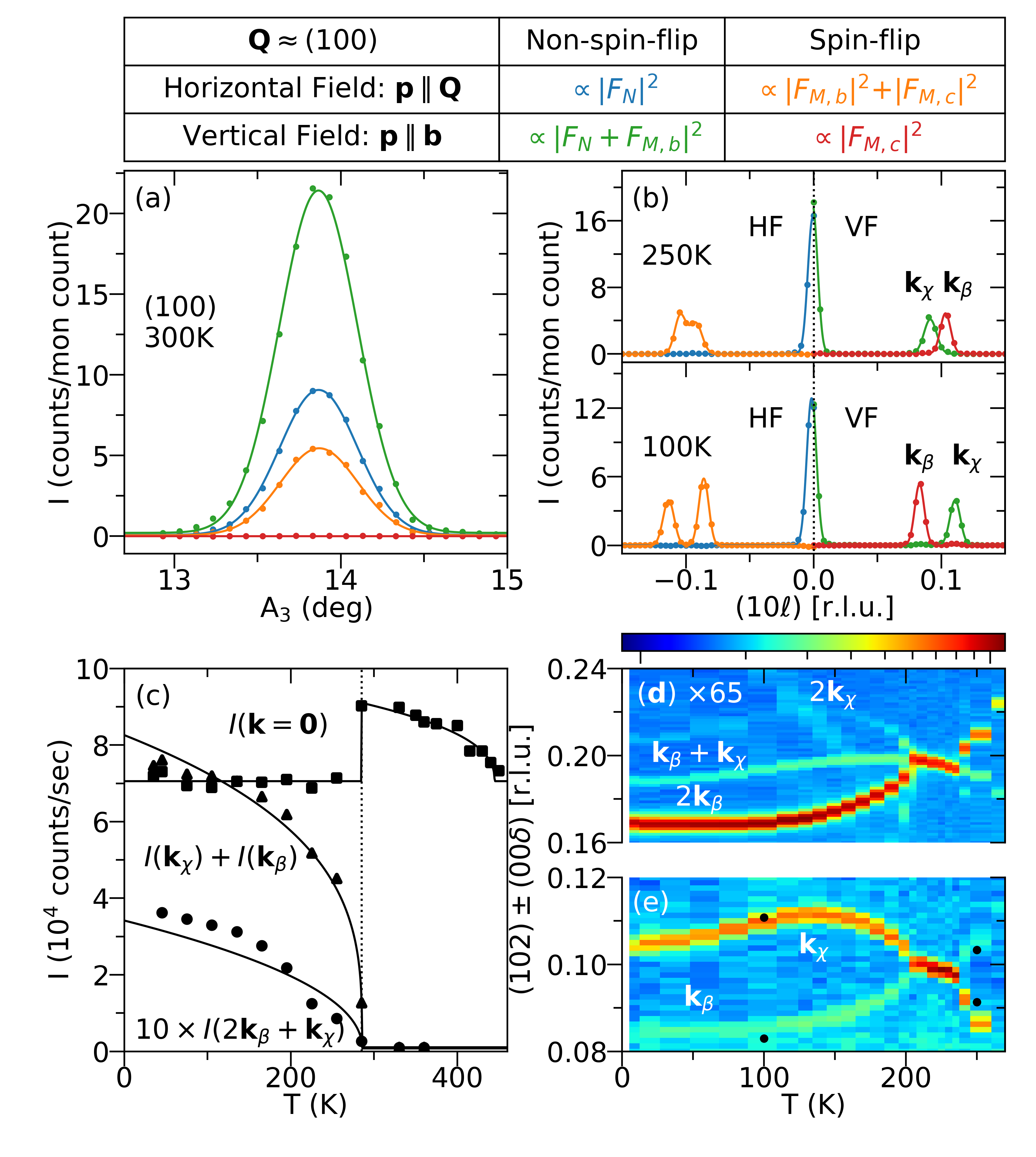}
    \centering
    \caption{Polarized beam data in this figure are color coded to indicate vertical polarization (VF) and horizontal polarization (HF), spin flip (SF) and non-spin-flip (NSF) data, probing magnetic and nuclear diffraction as indicated in the table above this figure. (a) Rocking scans through the $(100)$ Bragg peak of Mn$_3$Sn using polarized neutron diffraction at $T=300$~K. (b) Polarized neutron diffraction intensity versus wave vector transfer $\textbf{Q}=(10\ell)$ at  $T=250$~K and $T=100$~K. The positive and negative sides of the $(10\ell)$-axis show data for the VF and HF configuration, respectively. The solid lines in (a) and (b) are Gaussian fits.{ (c) Temperature dependence of the integrated unpolarized neutron scattering intensities  $I(\textbf{k}=\textbf{0}$), $I(\textbf{k}_{\chi}) + I(\textbf{k}_{\beta})$, and $I(2\textbf{k}_{\beta}+\textbf{k}_{\chi})$ acquired near the (100) Bragg peak. The solid lines show order parameter fits for each data set. Temperature dependence of (d) charge- and (e) magnetic-x-ray diffraction measured near the $(102)$ Bragg peak. A log scale is used and the scaling of the two vertical momentum axes in (d) and (e) differ by a factor 2 to emphasize the 2:1 ratio of the SDW versus CDW wavelengths. The black dots in (e) indicate the locations of the $k_{\beta}$ and $k_{\chi}$ peaks determined by polarized neutron diffraction.}}
    \label{fig: T_dependence}
\end{figure}

{We grew single crystals of Mn$_3$Sn with space group P6$_{3}$/mmc following the protocol described in the supplemental material (SM).}
To unambiguously resolve the character of each component of the complex order in Mn$_3$Sn, {we used neutrons polarized either along the momentum transfer $\bf Q$ (HF) or perpendicular to the scattering plane (VF). We resolved the spin-flip (SF) and the non-spin-flip (NSF) scattering for both configurations (see table in  Fig.~\ref{fig: T_dependence})}. Fig.~\ref{fig: T_dependence}(a) shows the intensity in each polarization channel for a rocking scan through the (100) Bragg peak at room temperature in the commensurate $\textbf{k~=~0}$ state. A comparison of the NSF channels (blue and green symbols) shows (100) is an allowed nuclear Bragg peak with magnetic scattering from spins polarized along the in-plane $\bf b$ direction. The lack of VF SF scattering (red symbols) shows the absence of c-polarized magnetism at 300 K. Fig.~\ref{fig: T_dependence}(b) shows scans along the $(10\ell)$ direction in the incommensurate phase for $T=250$~K and $T=100$~K. HF data (blue and orange symbols) are shown for $\ell<0$ and VF data (green and red symbols) for $\ell>0$. In contrast to Fig.~\ref{fig: T_dependence}(a), in the incommensurate phase the (100) peak has no SF intensity, which shows the $\textbf{k~=~0}$ magnetic order is absent. Instead magnetic diffraction is found for $|\ell|>0$, which indicates a spin structure that is modulated along  $c$. In the HF data, peaks are only present in the (orange) SF channel, which shows these arise from incommensurate magnetism (see table above Fig.~\ref{fig: T_dependence}). The VF data ($\ell>0$) show this magnetic order is non-co-planar with moments both within the basal plane (green) and along $\bf c$ (red), each with distinct temperature-dependent wave vectors that we denote by $\textbf{k}_{\chi}$ and $\textbf{k}_{\beta}$, respectively. 

The $T$-dependence of the neutron diffraction cross-section in Fig.~\ref{fig: T_dependence}(c) reveals two distinct phase transitions.  $\textbf{k~=~0}$ magnetic diffraction is superimposed upon nuclear diffraction for 285~K$<T<445$~K. The intensity grows continuously upon cooling below $T=445$~K as for a second-order phase transition. When the $\textbf{k~=~0}$ magnetic diffraction vanishes for $T<285$~K, the intensity of diffraction at $\bf k_\chi$ and $\bf k_\beta$ grows. We also see the onset of weak Bragg peaks with wave vector $2\textbf{k}_{\beta}+\textbf{k}_{\chi}$ (Fig.~\ref{fig: T_dependence}(c)). Contrary to a previous hypothesis~\cite{CABLE1993}, these data prove that the $\bf k_\chi$ and $\bf k_\beta$ magnetic components coexist within the same volume of Mn$_3$Sn.

To determine the temperature dependence of the wave vectors $\textbf{k}_{\chi}$ and $\textbf{k}_{\beta}$ with enhanced resolution,  we employed synchrotron x-ray diffraction. Magnetic diffraction patterns were acquired for $\bf Q$ along  $(1,0,2+\ell)$  and temperatures between 10~K and 270~K and are presented in Fig.~\ref{fig: T_dependence}(e). While $\textbf{k}_{\chi}$ and $\textbf{k}_{\beta}$ are well separated at the $T=285$~K onset of incommensurate magnetism, they approach each other upon cooling and merge into a single diffraction peak for 200~K$<T<230$~K. Upon cooling below 200~K, however, the peak splits again now with $k_{\chi}>k_{\beta}$. The peak assignment is accomplished by comparison to polarized magnetic neutron diffraction data (solid points in Fig.~\ref{fig: T_dependence}(e)). This complex interplay between the wave vector for the in-plane ($\textbf{k}_{\chi}$) and out-of-plane ($\textbf{k}_{\beta}$) polarized magnetic order affirms the homogeneous coexistence of these distinct components of the magnetic order. Upon further cooling, $\textbf{k}_{\beta}$ and $\textbf{k}_{\chi}$ stabilize to values that are indistinguishable from $\textbf{k}_{\beta} = \frac{1}{12}\textbf{c}^{*}$ and $\textbf{k}_{\chi} = \frac{5}{48}\textbf{c}^{*}$, respectively.


Fig.~\ref{fig: T_dependence}(d),  shows $2^{\rm nd}$ harmonic and interference peaks with wave vectors  $2\textbf{k}_{\chi}$, $2\textbf{k}_{\beta}$ and $\textbf{k}_{\chi}+\textbf{k}_{\beta}$, respectively. Because the analogous peaks are absent in our neutron diffraction measurements yet 2 orders of magnitude stronger than the magnetic x-ray diffraction peaks, we conclude they arise from a charge density modulation analogous to that found in Cr metal~\cite{RevModPhys.60.209}. The presence of the $\textbf{k}_{\chi}+\textbf{k}_{\beta}$ peak provides evidence of homogeneous coexistence of incommensurate SDW and CDW orders. For comparison $\rm CsV_3Sb_5$ displays CDW order~\cite{Ortiz2020} while FeGe develops a CDW within a collinear commensurate magnetic state\cite{teng2022,teng2023}.

\begin{figure}[t]
    \includegraphics[width=\columnwidth]{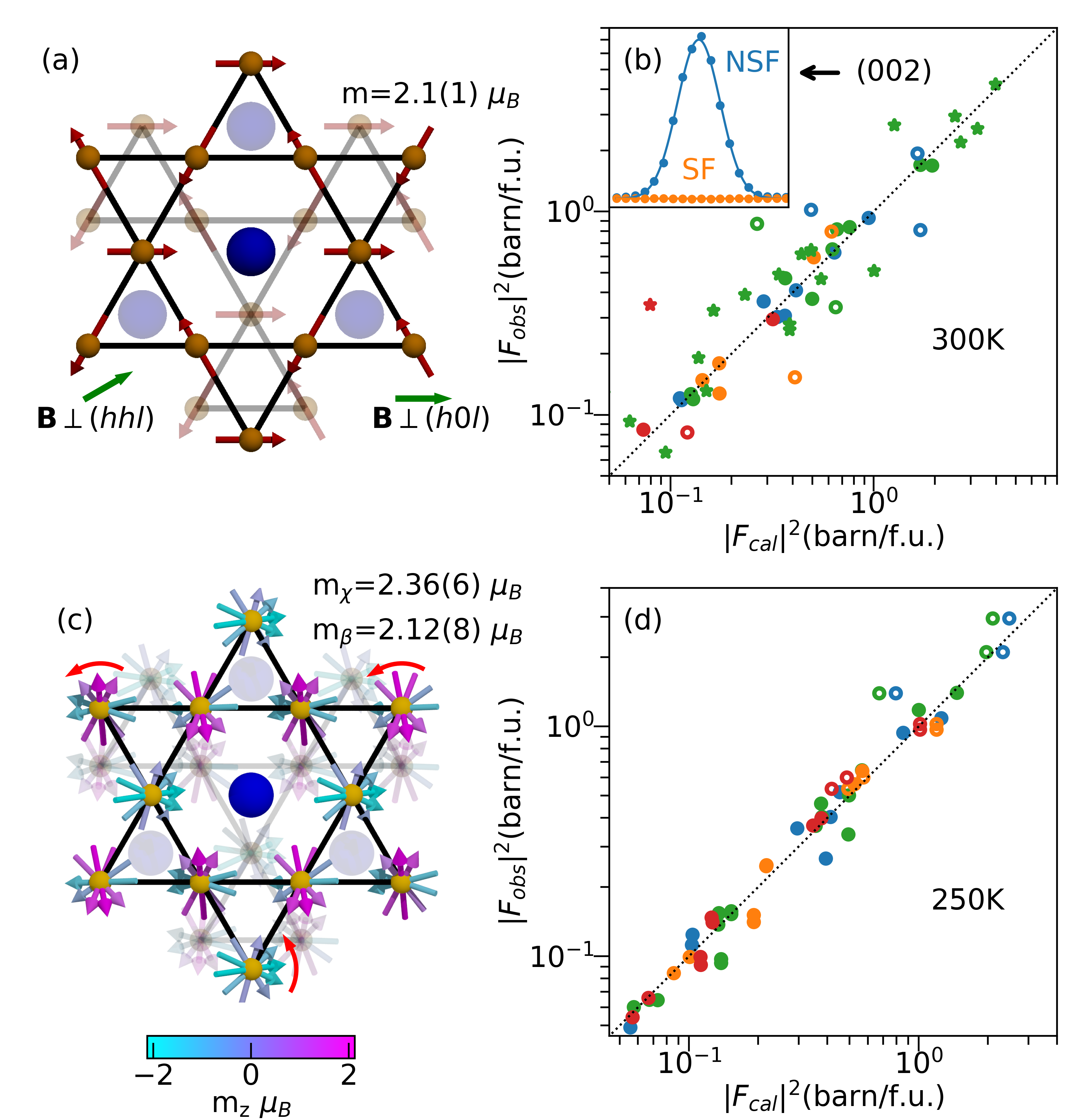}
    \centering
    \caption{Polarized beam data are color coded to indicate vertical polarization (VF) and horizontal polarization (HF), spin flip (SF) and non-spin-flip (NSF) data, probing magnetic and nuclear diffraction as indicated in the table above Fig.~\ref{fig: T_dependence}. (a) The $\textbf{k}~=~\textbf{0}$ magnetic structure of \ce{Mn3Sn} obtained at 300 K for the two indicated 2 T field directions. For the ${\bf B}\perp (hhl)$ field orientation only a second  symmetry related domain is anticipated. (b) Comparison between the observed and calculated structure factors for multi-domain (circles) and single domain 2 T measurements (stars). In frames (b) and (d) the four polarization channels are distinguished through the color scheme in the table above Fig.~\ref{fig: T_dependence} and open and closed symbols show data from the $(h0l)$ and $(hhl)$ planes, respectively. The inset shows the absence of ferromagnetic diffraction at $(002)$. (c) The proposed incommensurate structure where the color of each spin refers to the magnitude of its out-of-plane component. The curved red arrows indicate how the direction of the spins is rotated between adjacent kagome planes. (d) Comparison between the observed and calculated structure factors for the incommensurate phase at 250~K.}
    \label{fig: results}
\end{figure}

For detailed quantitative information about the commensurate and incommensurate phases of \ce{Mn3Sn}, we refine the polarization-resolved intensity of scans along ${\bf c}^*$ through magnetic peaks accessible with 14.7 meV neutrons in the $(h0\ell)$ and $(hh\ell)$ scattering plane. For the $\textbf{k}=\textbf{0}$ commensurate phase, which exists for $ T\in [285,445]$~K (Fig.~\ref{fig: T_dependence}(c)), we collected room temperature rocking scans for all four polarization channels at the accessible $\textbf{k}=\textbf{0}$ Bragg peaks. The second-order critical behavior indicated by Fig.~\ref{fig: T_dependence}(c) ensures the $\textbf{k}=\textbf{0}$ spin structure is associated with a single irreducible representation (IR). The only IR that allows for the absence of magnetic diffraction at $(111)$ and the existence of a $(110)$ peak within the SF and NSF channels of all polarization configurations is $\Gamma_9$~\cite{PhysRevB.102.054403,Brown_1990}. The absence of magnetic diffraction at $\textbf{Q}=(002)$ (inset of Fig.~\ref{fig: results}(b)) allows a description in terms of a pure anti-chiral antiferromagnetic spin structure as the small uniform magnetization of 0.007~$\mu_B$/Mn\cite{Nakatsuji2015} does not produce significant diffraction. Thus, the parameters to be refined within $\Gamma_9$ are the moment size ($M_{\chi}$) and a uniform rotation of all spins about the c-axis. However, in a multi-domain sample with a macroscopic 6-fold axis, even polarized neutron diffraction is insensitive to this angular variable~\cite{Brown_1990}. We therefore acquired polarized neutron diffraction data in a 2 T field, which exceeds the coercive field inferred from magnetization data. The field was applied along ${\bf b}$  and  $[1\bar{1}0]$, which are perpendicular to the $(h0\ell)$ and $(hh\ell)$ scattering planes, respectively as shown in Fig.~\ref{fig: results}(a). The quality of the 0~T and the 2~T refinements can be gauged from Fig.~\ref{fig: results}(b). The best fit  is obtained with $M_{\chi}=2.1(1)\mu_{B}$/Mn and all spins parallel to the edges of the equilateral triangles that form the kagome lattices as shown with each of the two applied field directions in Fig.~\ref{fig: results}(a). This shows that $\langle 100\rangle$ directions are easy axes for Mn$_3$Sn as opposed to the $\langle 1\bar{1}0 \rangle$ directions for Mn$_3$Ge\cite{PhysRevB.101.140411}. 

To determine the low temperature modulated spin structure of \ce{Mn3Sn}, we acquired polarized diffraction data at $T=250$~K.
HF, VF, NSF and SF Bragg diffraction cross sections were obtained by integrating the corresponding intensity of $(h0\ell)$ and $(hh\ell)$ scans holding $h$ constant through magnetic Bragg peaks of the form $\textbf{G}\pm\textbf{k}_{\chi}$ and $\textbf{G}\pm\textbf{k}_{\beta}$ using 14.7 meV neutrons. Here $\textbf{G}$ is a nuclear Bragg peak position, $\textbf{k}_{\chi}=(0,0,\pm0.092(1))$, and $\textbf{k}_{\beta}=(0,0,\pm0.104(1))$. The data can be described by the $\Gamma_{6}$ IR of the 'little group' $G_{k_{\chi/\beta}}$. The 6 basis vectors of $\Gamma_{6}$ are defined in the supplementary material. $\chi_{x}$, $\chi_{y}$, $f_{x}$, and $f_{y}$ are amplitudes of kagome plane components while  $\beta_{1}$ and $\beta_{2}$ describe the out-of-plane spin modulation. The detailed structures for IRs are included in Supplementary material. For $\textbf{G}$~=~(100), (200), (300), and (110), the $\textbf{k}_{\chi}$ and the $\textbf{k}_{\beta}$ peaks respectively appear in the NSF and SF channels of the VF configuration (Fig.~\ref{fig: T_dependence}(b)). Thus the $\textbf{k}_{\chi}$ component of the spin structure is associated with $\chi_{x}$, $\chi_{y}$, $f_{x}$, $f_{y}$ while the $\textbf{k}_{\beta}$ component must be described by $\beta_{1}$ and $\beta_{2}$. Considering the multi-domain nature of the order, we obtain the following constraints on the amplitudes of the basis vectors at $T=250$~K: 
\begin{eqnarray}
m_\chi^2=2(|\chi_{x}|^{2}+|\chi_{y}|^{2})&=&(2.36(6)\mu_B)^{2}\\
m_\beta^2=2(|\beta_{1}|^{2}+|\beta_{2}|^{2})&=&(2.12(8)\mu_B)^{2}\\
|f_{x}|=|f_{y}|&=&0.00(7)\mu_B.
\end{eqnarray} 
A spin structure consistent with this polarized neutron diffraction refinement is depicted in Fig.~\ref{fig: results}(c). For this structure, both $\chi_{x}$ and $\beta_{1}$ are real while $\chi_{y}~=~i\chi_{x}$ and $\beta_{2}~=~i\beta_{1}$. This yields a helical in-plane anti-chiral order with a moment of $2.36(6)\mu_{B}$/Mn superimposed on an out-of-plane modulation with an amplitude of 2.12(8)$\mu_{B}$/Mn. A comparison between the $T=250$~K observed and calculated neutron structure factors is in Fig.~\ref{fig: results}(d). 

\begin{figure}[t]
    \includegraphics[width=\columnwidth]{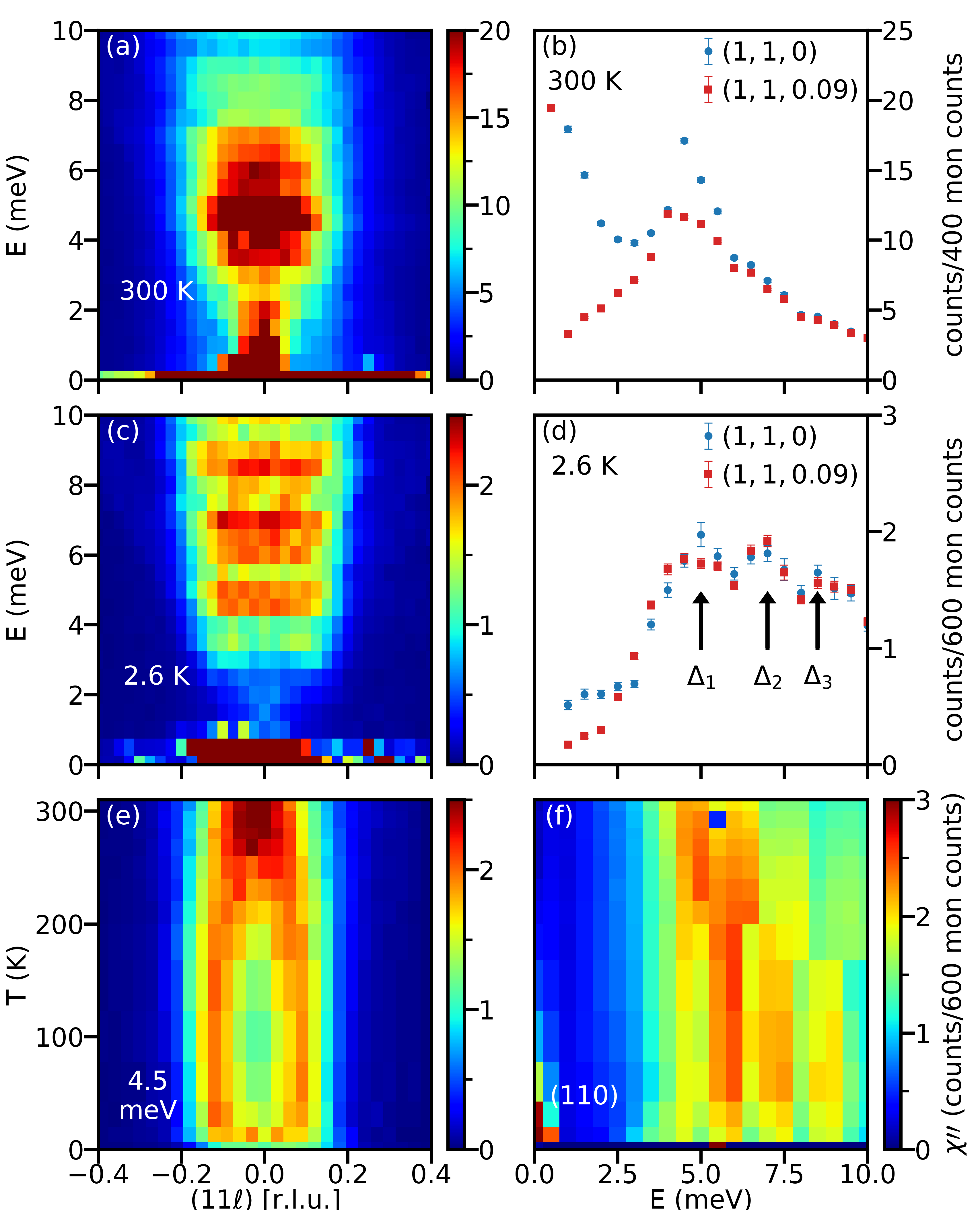}
\caption{(a) The (00$\ell$)-dependence of the 300~K inelastic neutron scattering from \ce{Mn3Sn} acquired near (110). (b) Constant {\bf{Q}}-cuts at $(110)$ and $(1,1,\pm 0.09)$ obtained from the 300 K data  in (a). (c,d) Inelastic neutron scattering acquired at 2.6~K and presented as for panels (a,b).  (e) The temperature dependence of the dynamic susceptibility $\chi^{\prime\prime}({\bf Q}=(11\ell),\hbar\omega=4.5$~meV). (f) The temperature dependence of the $\ell\in [-0.25,0.25]$-integrated local dynamic susceptibility shows the appearance of two additional modes $\Delta_{2}$ and $\Delta_{3}$ for $T<285$~K in the incommensurate phase.}
    \label{fig: spectra}
\end{figure}

\begin{figure}[t]
    \includegraphics[width=\columnwidth]{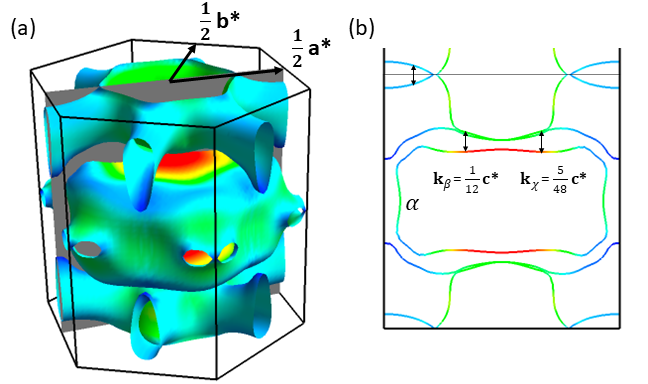}
    \centering
    \caption{Density functional theory calculation for Mn$_3$Sn in the anti-chiral $\mathbf{k}= \mathbf{0}$ phase. (a) Fermi surfaces where flat sheets extending within the Kagome plane admit nesting for small wave vectors extending along the {\bf c} axis. (b) shows the slice indicated by the gray surface in (a) through the Fermi surfaces. The potential nesting wave vectors are labeled in panel (b), including a interzone nesting. The surface marked  $\alpha$,  contains Weyl point in the $\mathbf{k}= \mathbf{0}$ phase and is nested with an enclosing surface by the observed magnetic wave vectors $\mathbf{k}_{\chi,\beta}$.}
    \label{fig: DFT}
\end{figure}

To understand the origin of this non-coplanar modulated phase, we examine the associated soft modes through cold neutron spectroscopy with wave vector transfer near (110). Fig.~\ref{fig: spectra}(a,b) show the excitation spectrum in the anti-chiral commensurate phase for wave vector transfer $\mathbf{Q}=(11\ell), |\ell|<0.4$. As for Mn$_3$Ge, the spectrum has a gapless mode visible only very close to (110) and a 4.5 meV resonance that extends to wave vector transfer $|\ell|<0.15$\cite{PhysRevB.102.054403}. The presence of the resonance in THz Faraday rotation spectroscopy on $\rm Mn_{3+x}Sn_{1-x}$ ($x=0.13, 0.47$) thin films\cite{Khadkaeabc1977} confirms it has spectral weight at the zone center. We associate the gapless mode with the in-plane Goldstone mode that arises because the anti-chiral magnetic order breaks the in-kagome-plane rotational degree of freedom for Mn-spins  (Fig.~\ref{fig: results}(a)). There may also be contributions to this scattering from acoustic phonons. We associate the 4.5 meV resonance with a doublet of out-of-plane polarized spin waves $\beta_1$ and $\beta_2$, which are degenerate when the in-plane anisotropy is negligible\cite{PhysRevB.102.144417}. The co-planar anti-chiral magnetic structure of the $\mathbf{k}=\textbf{0}$ phase in \ce{Mn3X} can be described by localized magnetic moments with nearest-neighbor antiferromagnetic and Dzyaloshinskii-Moriya (DM) interactions. The DM vectors of the in-plane nearest neighbor interactions are pinned along the $\mathbf{c}$ axis by crystal symmetry and determine the chirality of the antiferromagnetic structure.

Upon cooling into the incommensurate state, there is a chromium-like increase in the c-axis resistivity \cite{doi:10.7566/JPSCP.30.011177}, which indicates the opening of gaps on nested parts of the Fermi surface. While the co-planar anti-chiral order converts to a transverse spiral with wave vector ${\bf k}_\chi$, its amplitude is virtually unchanged across the transition (Fig.~\ref{fig: results}(a,c)). This suggests that out-of-plane rather than in-plane spin fluctuations condense at the incommensurate phase transition so that spectral weight is drawn from the $\beta_1$ and $\beta_2$ modes at 4.5 meV to form the $\mathbf{k}_{\beta}$ Bragg peaks. Thus it appears that ${\bf c}$-polarized SDW with wave vector ${\bf k}_\beta$ along with a CDW at $2{\bf k}_\beta$ are the associated amplitude-modulated electronic density waves. 

In the modulated non-coplanar phase of \ce{Mn3Sn} at 2.6~K, Figs.~\ref{fig: spectra}(d,f) show three dominant spectral components at $\Delta_{1}=5.0(5)~\mathrm{meV},\Delta_{2}=7.0(5)~\mathrm{meV}$ and $\Delta_{3}=8.5(5)~\mathrm{meV}$. While there are precisely three low energy modes for an equilateral triangle of antiferromagnetically interacting spins\cite{PhysRevB.102.144417}, the multi-${\bf k}$ long-wavelength nature of the structure may contribute to the complexity of the low $T$ spectrum.  Fig.~\ref{fig: spectra}(e) shows that the scattering intensity at $\hbar\omega=4.5$~meV continuously shifts from $\mathbf{k}=\textbf{0}$ to $\mathbf{k}=\mathbf{k_{\chi,\beta}}$ {upon cooling. For $T<50$~K the $\mathbf{k}\approx{\mathbf{0}}$ region between $\mathbf{k_{\chi,\beta}}$ peaks regains intensity. Though the static magnetism remains modulated, this is the temperature range where all wave vectors become temperature-independent and indistinguishable from rational fractions (Fig.~\ref{fig: T_dependence}(d,e)). The return to commensurability and a finite set of local environments for Mn may play a role in the modified $\Gamma$-point spectrum as well as in the $T_g=31.5$~K spin-glass transition\cite{PhysRevB.73.205105}.}

To examine whether Fermi-surface nesting precipitates these developments, the density functional electronic band structure was calculated. The corresponding Fermi surfaces shown in Fig.~\ref{fig: DFT} contain large flat sheets extending perpendicular to the c-axis. As shown in Fig.~\ref{fig: DFT}(b), large areas of these Fermi-surfaces are connected by  ${\bf k}_\beta$ and ${\bf k}_\chi$. {Nesting is sensitive to small changes in the chemical potential, which may explain the sample-dependent incommensurability of Mn$_3$Sn~\cite{CABLE1993}.} Furthermore, one of the nested Fermi surfaces marked $\alpha$ in Fig.~\ref{fig: DFT}(b) contains Weyl nodes at room temperature~\cite{Kuroda2017,chen2021anomalous} so that a change in band topology can be anticipated as the bands hybridize to form spin and charge density waves. This is indicated by the suppression of both the anomalous Hall and Nernst effects~\cite{PhysRevLett.119.056601,doi:10.1063/1.5021133} as required by the fact that the incommensurate phase with $k_\chi\neq 0$ cannot define a direction within the basal plane. The different temperature dependence of spatial modulations observed for $\mathbf{k}_{\beta}$ and $\mathbf{k}_{\chi}$ resembles observations in the kagome semi-metal YMn$_6$Sn$_6$\cite{doi:10.1126/sciadv.abe2680}, which may reflect the complex flat band nature of the nesting.

While aspects of the low energy magnetism in the commensurate phases of Mn$_3$Sn\cite{Park2018} and Mn$_3$Ge\cite{PhysRevB.102.054403, PhysRevB.102.144417} can be described by models of interacting local moments\cite{PhysRevLett.119.087202}, the incommensurate CDW and SDW phases of Mn$_3$Sn reported here manifest a complex interplay between conduction electrons, magnetism and band topology in topological kagome materials, revealing a new form of order peculiar to a correlated topological state. Our work lays foundations for developing understanding of the role of correlation in topological magnets and its applications for spintronics and thermoelectronics~\cite{doi:10.1146/annurev-conmatphys-031620-103859}

\begin{acknowledgments}
This work was supported as part of the Institute for Quantum Matter, an Energy Frontier Research Center funded by the U.S. Department of Energy, Office of Science, Basic Energy Sciences under Award No. DE-SC0019331. J.G. acknowledges support from the NSERC Postdoctoral Fellowship Program and C.B. acknowledges support from the Gordon and Betty Moore Foundation GBMF9456. Work at the University of Tokyo was supported by JST-Mirai Program (JPMJMI20A1), JST-CREST (JPMJCR18T3), JST-ASPIRE (JPMJAP2317). A portion of this research used resources at the High Flux Isotope Reactor and Spallation Neutron Source, a DOE Office of Science User Facility operated by the Oak Ridge National Laboratory. Access to MACS was provided by the Center for High-Resolution Neutron Scattering, a partnership between the National Institute of Standards and Technology and the National Science Foundation under Agreement No. DMR-1508249. The X-ray diffraction experiments used resources from the Advanced Photon Source, a U.S. DOE Office of Science User Facility operated for the DOE Office of Science by Argonne National Laboratory under Contract No. DE-AC02-06CH11357.
\end{acknowledgments}
\noindent
\bibliography{Mn3Sn}

\begin{thebibliography}{38}%
\makeatletter
\providecommand \@ifxundefined [1]{%
 \@ifx{#1\undefined}
}%
\providecommand \@ifnum [1]{%
 \ifnum #1\expandafter \@firstoftwo
 \else \expandafter \@secondoftwo
 \fi
}%
\providecommand \@ifx [1]{%
 \ifx #1\expandafter \@firstoftwo
 \else \expandafter \@secondoftwo
 \fi
}%
\providecommand \natexlab [1]{#1}%
\providecommand \enquote  [1]{``#1''}%
\providecommand \bibnamefont  [1]{#1}%
\providecommand \bibfnamefont [1]{#1}%
\providecommand \citenamefont [1]{#1}%
\providecommand \href@noop [0]{\@secondoftwo}%
\providecommand \href [0]{\begingroup \@sanitize@url \@href}%
\providecommand \@href[1]{\@@startlink{#1}\@@href}%
\providecommand \@@href[1]{\endgroup#1\@@endlink}%
\providecommand \@sanitize@url [0]{\catcode `\\12\catcode `\$12\catcode
  `\&12\catcode `\#12\catcode `\^12\catcode `\_12\catcode `\%12\relax}%
\providecommand \@@startlink[1]{}%
\providecommand \@@endlink[0]{}%
\providecommand \url  [0]{\begingroup\@sanitize@url \@url }%
\providecommand \@url [1]{\endgroup\@href {#1}{\urlprefix }}%
\providecommand \urlprefix  [0]{URL }%
\providecommand \Eprint [0]{\href }%
\providecommand \doibase [0]{http://dx.doi.org/}%
\providecommand \selectlanguage [0]{\@gobble}%
\providecommand \bibinfo  [0]{\@secondoftwo}%
\providecommand \bibfield  [0]{\@secondoftwo}%
\providecommand \translation [1]{[#1]}%
\providecommand \BibitemOpen [0]{}%
\providecommand \bibitemStop [0]{}%
\providecommand \bibitemNoStop [0]{.\EOS\space}%
\providecommand \EOS [0]{\spacefactor3000\relax}%
\providecommand \BibitemShut  [1]{\csname bibitem#1\endcsname}%
\let\auto@bib@innerbib\@empty
\bibitem [{\citenamefont {Liu}\ \emph {et~al.}(2018)\citenamefont {Liu},
  \citenamefont {Sun}, \citenamefont {Kumar}, \citenamefont {Muechler},
  \citenamefont {Sun}, \citenamefont {Jiao}, \citenamefont {Yang},
  \citenamefont {Liu}, \citenamefont {Liang}, \citenamefont {Xu}, \citenamefont
  {Kroder}, \citenamefont {S\"u\ss}, \citenamefont {Borrmann}, \citenamefont
  {Shekhar}, \citenamefont {Wang}, \citenamefont {Xi}, \citenamefont {Wang},
  \citenamefont {Schnelle}, \citenamefont {Wirth}, \citenamefont {Chen},
  \citenamefont {Goennenwein},\ and\ \citenamefont {Felser}}]{liu2018giant}%
  \BibitemOpen
  \bibfield  {author} {\bibinfo {author} {\bibfnamefont {E.}~\bibnamefont
  {Liu}}, \bibinfo {author} {\bibfnamefont {Y.}~\bibnamefont {Sun}}, \bibinfo
  {author} {\bibfnamefont {N.}~\bibnamefont {Kumar}}, \bibinfo {author}
  {\bibfnamefont {L.}~\bibnamefont {Muechler}}, \bibinfo {author}
  {\bibfnamefont {A.}~\bibnamefont {Sun}}, \bibinfo {author} {\bibfnamefont
  {L.}~\bibnamefont {Jiao}}, \bibinfo {author} {\bibfnamefont {S.-Y.}\
  \bibnamefont {Yang}}, \bibinfo {author} {\bibfnamefont {D.}~\bibnamefont
  {Liu}}, \bibinfo {author} {\bibfnamefont {A.}~\bibnamefont {Liang}}, \bibinfo
  {author} {\bibfnamefont {Q.}~\bibnamefont {Xu}}, \bibinfo {author}
  {\bibfnamefont {J.}~\bibnamefont {Kroder}}, \bibinfo {author} {\bibfnamefont
  {V.}~\bibnamefont {S\"u\ss}}, \bibinfo {author} {\bibfnamefont
  {H.}~\bibnamefont {Borrmann}}, \bibinfo {author} {\bibfnamefont
  {C.}~\bibnamefont {Shekhar}}, \bibinfo {author} {\bibfnamefont
  {Z.}~\bibnamefont {Wang}}, \bibinfo {author} {\bibfnamefont {C.}~\bibnamefont
  {Xi}}, \bibinfo {author} {\bibfnamefont {W.}~\bibnamefont {Wang}}, \bibinfo
  {author} {\bibfnamefont {W.}~\bibnamefont {Schnelle}}, \bibinfo {author}
  {\bibfnamefont {S.}~\bibnamefont {Wirth}}, \bibinfo {author} {\bibfnamefont
  {Y.}~\bibnamefont {Chen}}, \bibinfo {author} {\bibfnamefont {S.~T.~B.}\
  \bibnamefont {Goennenwein}}, \ and\ \bibinfo {author} {\bibfnamefont
  {C.}~\bibnamefont {Felser}},\ }\href {\doibase
  https://doi.org/10.1038/s41567-018-0234-5} {\bibfield  {journal} {\bibinfo
  {journal} {Nat. Phys.}\ }\textbf {\bibinfo {volume} {14}},\ \bibinfo {pages}
  {1125} (\bibinfo {year} {2018})}\BibitemShut {NoStop}%
\bibitem [{\citenamefont {Nakatsuji}\ \emph {et~al.}(2015)\citenamefont
  {Nakatsuji}, \citenamefont {Kiyohara},\ and\ \citenamefont
  {Higo}}]{Nakatsuji2015}%
  \BibitemOpen
  \bibfield  {author} {\bibinfo {author} {\bibfnamefont {S.}~\bibnamefont
  {Nakatsuji}}, \bibinfo {author} {\bibfnamefont {N.}~\bibnamefont {Kiyohara}},
  \ and\ \bibinfo {author} {\bibfnamefont {T.}~\bibnamefont {Higo}},\ }\href
  {\doibase 10.1038/nature15723} {\bibfield  {journal} {\bibinfo  {journal}
  {Nature}\ }\textbf {\bibinfo {volume} {527}},\ \bibinfo {pages} {212}
  (\bibinfo {year} {2015})}\BibitemShut {NoStop}%
\bibitem [{\citenamefont {Armitage}\ \emph {et~al.}(2018)\citenamefont
  {Armitage}, \citenamefont {Mele},\ and\ \citenamefont
  {Vishwanath}}]{Armitage2018}%
  \BibitemOpen
  \bibfield  {author} {\bibinfo {author} {\bibfnamefont {N.~P.}\ \bibnamefont
  {Armitage}}, \bibinfo {author} {\bibfnamefont {E.~J.}\ \bibnamefont {Mele}},
  \ and\ \bibinfo {author} {\bibfnamefont {A.}~\bibnamefont {Vishwanath}},\
  }\href {\doibase https://doi.org/10.1103/RevModPhys.90.015001} {\bibfield
  {journal} {\bibinfo  {journal} {Rev. Mod. Phys.}\ }\textbf {\bibinfo {volume}
  {90}},\ \bibinfo {pages} {015001} (\bibinfo {year} {2018})}\BibitemShut
  {NoStop}%
\bibitem [{\citenamefont {Ortiz}\ \emph {et~al.}(2019)\citenamefont {Ortiz},
  \citenamefont {Gomes}, \citenamefont {Morey}, \citenamefont {Winiarski},
  \citenamefont {Bordelon}, \citenamefont {Mangum}, \citenamefont {Oswald},
  \citenamefont {Rodriguez-Rivera}, \citenamefont {Neilson}, \citenamefont
  {Wilson}, \citenamefont {Ertekin}, \citenamefont {McQueen},\ and\
  \citenamefont {Toberer}}]{WOS:000486663000001}%
  \BibitemOpen
  \bibfield  {author} {\bibinfo {author} {\bibfnamefont {B.~R.}\ \bibnamefont
  {Ortiz}}, \bibinfo {author} {\bibfnamefont {L.~C.}\ \bibnamefont {Gomes}},
  \bibinfo {author} {\bibfnamefont {J.~R.}\ \bibnamefont {Morey}}, \bibinfo
  {author} {\bibfnamefont {M.}~\bibnamefont {Winiarski}}, \bibinfo {author}
  {\bibfnamefont {M.}~\bibnamefont {Bordelon}}, \bibinfo {author}
  {\bibfnamefont {J.~S.}\ \bibnamefont {Mangum}}, \bibinfo {author}
  {\bibfnamefont {L.~W.~H.}\ \bibnamefont {Oswald}}, \bibinfo {author}
  {\bibfnamefont {J.~A.}\ \bibnamefont {Rodriguez-Rivera}}, \bibinfo {author}
  {\bibfnamefont {J.~R.}\ \bibnamefont {Neilson}}, \bibinfo {author}
  {\bibfnamefont {S.~D.}\ \bibnamefont {Wilson}}, \bibinfo {author}
  {\bibfnamefont {E.}~\bibnamefont {Ertekin}}, \bibinfo {author} {\bibfnamefont
  {T.~M.}\ \bibnamefont {McQueen}}, \ and\ \bibinfo {author} {\bibfnamefont
  {E.~S.}\ \bibnamefont {Toberer}},\ }\href {\doibase
  10.1103/PhysRevMaterials.3.094407} {\bibfield  {journal} {\bibinfo  {journal}
  {Physical Review Materials}\ }\textbf {\bibinfo {volume} {3}} (\bibinfo
  {year} {2019}),\ 10.1103/PhysRevMaterials.3.094407}\BibitemShut {NoStop}%
\bibitem [{\citenamefont {Jiang}\ \emph {et~al.}(2021)\citenamefont {Jiang},
  \citenamefont {Yin}, \citenamefont {Denner}, \citenamefont {Shumiya},
  \citenamefont {Ortiz}, \citenamefont {Xu}, \citenamefont {Guguchia},
  \citenamefont {He}, \citenamefont {Hossain}, \citenamefont {Liu},
  \citenamefont {Ruff}, \citenamefont {Kautzsch}, \citenamefont {Zhang},
  \citenamefont {Chang}, \citenamefont {Belopolski}, \citenamefont {Zhang},
  \citenamefont {Cochran}, \citenamefont {Multer}, \citenamefont {Litskevich},
  \citenamefont {Cheng}, \citenamefont {Yang}, \citenamefont {Wang},
  \citenamefont {Thomale}, \citenamefont {Neupert}, \citenamefont {Wilson},\
  and\ \citenamefont {Hasan}}]{Jiang2021}%
  \BibitemOpen
  \bibfield  {author} {\bibinfo {author} {\bibfnamefont {Y.-X.}\ \bibnamefont
  {Jiang}}, \bibinfo {author} {\bibfnamefont {J.-X.}\ \bibnamefont {Yin}},
  \bibinfo {author} {\bibfnamefont {M.~M.}\ \bibnamefont {Denner}}, \bibinfo
  {author} {\bibfnamefont {N.}~\bibnamefont {Shumiya}}, \bibinfo {author}
  {\bibfnamefont {B.~R.}\ \bibnamefont {Ortiz}}, \bibinfo {author}
  {\bibfnamefont {G.}~\bibnamefont {Xu}}, \bibinfo {author} {\bibfnamefont
  {Z.}~\bibnamefont {Guguchia}}, \bibinfo {author} {\bibfnamefont
  {J.}~\bibnamefont {He}}, \bibinfo {author} {\bibfnamefont {M.~S.}\
  \bibnamefont {Hossain}}, \bibinfo {author} {\bibfnamefont {X.}~\bibnamefont
  {Liu}}, \bibinfo {author} {\bibfnamefont {J.}~\bibnamefont {Ruff}}, \bibinfo
  {author} {\bibfnamefont {L.}~\bibnamefont {Kautzsch}}, \bibinfo {author}
  {\bibfnamefont {S.~S.}\ \bibnamefont {Zhang}}, \bibinfo {author}
  {\bibfnamefont {G.}~\bibnamefont {Chang}}, \bibinfo {author} {\bibfnamefont
  {I.}~\bibnamefont {Belopolski}}, \bibinfo {author} {\bibfnamefont
  {Q.}~\bibnamefont {Zhang}}, \bibinfo {author} {\bibfnamefont {T.~A.}\
  \bibnamefont {Cochran}}, \bibinfo {author} {\bibfnamefont {D.}~\bibnamefont
  {Multer}}, \bibinfo {author} {\bibfnamefont {M.}~\bibnamefont {Litskevich}},
  \bibinfo {author} {\bibfnamefont {Z.-J.}\ \bibnamefont {Cheng}}, \bibinfo
  {author} {\bibfnamefont {X.~P.}\ \bibnamefont {Yang}}, \bibinfo {author}
  {\bibfnamefont {Z.}~\bibnamefont {Wang}}, \bibinfo {author} {\bibfnamefont
  {R.}~\bibnamefont {Thomale}}, \bibinfo {author} {\bibfnamefont
  {T.}~\bibnamefont {Neupert}}, \bibinfo {author} {\bibfnamefont {S.~D.}\
  \bibnamefont {Wilson}}, \ and\ \bibinfo {author} {\bibfnamefont {M.~Z.}\
  \bibnamefont {Hasan}},\ }\href {\doibase 10.1038/s41563-021-01034-y}
  {\bibfield  {journal} {\bibinfo  {journal} {Nature Materials}\ }\textbf
  {\bibinfo {volume} {20}},\ \bibinfo {pages} {1353} (\bibinfo {year}
  {2021})}\BibitemShut {NoStop}%
\bibitem [{\citenamefont {Ortiz}\ \emph {et~al.}(2021)\citenamefont {Ortiz},
  \citenamefont {Sarte}, \citenamefont {Kenney}, \citenamefont {Graf},
  \citenamefont {Teicher}, \citenamefont {Seshadri},\ and\ \citenamefont
  {Wilson}}]{PhysRevMaterials.5.034801}%
  \BibitemOpen
  \bibfield  {author} {\bibinfo {author} {\bibfnamefont {B.~R.}\ \bibnamefont
  {Ortiz}}, \bibinfo {author} {\bibfnamefont {P.~M.}\ \bibnamefont {Sarte}},
  \bibinfo {author} {\bibfnamefont {E.~M.}\ \bibnamefont {Kenney}}, \bibinfo
  {author} {\bibfnamefont {M.~J.}\ \bibnamefont {Graf}}, \bibinfo {author}
  {\bibfnamefont {S.~M.~L.}\ \bibnamefont {Teicher}}, \bibinfo {author}
  {\bibfnamefont {R.}~\bibnamefont {Seshadri}}, \ and\ \bibinfo {author}
  {\bibfnamefont {S.~D.}\ \bibnamefont {Wilson}},\ }\href {\doibase
  10.1103/PhysRevMaterials.5.034801} {\bibfield  {journal} {\bibinfo  {journal}
  {Phys. Rev. Mater.}\ }\textbf {\bibinfo {volume} {5}},\ \bibinfo {pages}
  {034801} (\bibinfo {year} {2021})}\BibitemShut {NoStop}%
\bibitem [{\citenamefont {Ortiz}\ \emph
  {et~al.}(2020{\natexlab{a}})\citenamefont {Ortiz}, \citenamefont {Teicher},
  \citenamefont {Hu}, \citenamefont {Zuo}, \citenamefont {Sarte}, \citenamefont
  {Schueller}, \citenamefont {Abeykoon}, \citenamefont {Krogstad},
  \citenamefont {Rosenkranz}, \citenamefont {Osborn}, \citenamefont {Seshadri},
  \citenamefont {Balents}, \citenamefont {He},\ and\ \citenamefont
  {Wilson}}]{ortiz2020cs}%
  \BibitemOpen
  \bibfield  {author} {\bibinfo {author} {\bibfnamefont {B.~R.}\ \bibnamefont
  {Ortiz}}, \bibinfo {author} {\bibfnamefont {S.~M.~L.}\ \bibnamefont
  {Teicher}}, \bibinfo {author} {\bibfnamefont {Y.}~\bibnamefont {Hu}},
  \bibinfo {author} {\bibfnamefont {J.~L.}\ \bibnamefont {Zuo}}, \bibinfo
  {author} {\bibfnamefont {P.~M.}\ \bibnamefont {Sarte}}, \bibinfo {author}
  {\bibfnamefont {E.~C.}\ \bibnamefont {Schueller}}, \bibinfo {author}
  {\bibfnamefont {A.~M.~M.}\ \bibnamefont {Abeykoon}}, \bibinfo {author}
  {\bibfnamefont {M.~J.}\ \bibnamefont {Krogstad}}, \bibinfo {author}
  {\bibfnamefont {S.}~\bibnamefont {Rosenkranz}}, \bibinfo {author}
  {\bibfnamefont {R.}~\bibnamefont {Osborn}}, \bibinfo {author} {\bibfnamefont
  {R.}~\bibnamefont {Seshadri}}, \bibinfo {author} {\bibfnamefont
  {L.}~\bibnamefont {Balents}}, \bibinfo {author} {\bibfnamefont
  {J.}~\bibnamefont {He}}, \ and\ \bibinfo {author} {\bibfnamefont {S.~D.}\
  \bibnamefont {Wilson}},\ }\href {\doibase
  https://doi.org/10.1103/PhysRevLett.125.247002} {\bibfield  {journal}
  {\bibinfo  {journal} {Phys. Rev. Lett.}\ }\textbf {\bibinfo {volume} {125}},\
  \bibinfo {pages} {247002} (\bibinfo {year} {2020}{\natexlab{a}})}\BibitemShut
  {NoStop}%
\bibitem [{\citenamefont {Li}\ \emph {et~al.}(2021)\citenamefont {Li},
  \citenamefont {Zhang}, \citenamefont {Yilmaz}, \citenamefont {Pai},
  \citenamefont {Marvinney}, \citenamefont {Said}, \citenamefont {Yin},
  \citenamefont {Gong}, \citenamefont {Tu}, \citenamefont {Vescovo},
  \citenamefont {Nelson}, \citenamefont {Moore}, \citenamefont {Murakami},
  \citenamefont {Lei}, \citenamefont {Lee}, \citenamefont {Lawrie},\ and\
  \citenamefont {Miao}}]{PhysRevX.11.031050}%
  \BibitemOpen
  \bibfield  {author} {\bibinfo {author} {\bibfnamefont {H.}~\bibnamefont
  {Li}}, \bibinfo {author} {\bibfnamefont {T.~T.}\ \bibnamefont {Zhang}},
  \bibinfo {author} {\bibfnamefont {T.}~\bibnamefont {Yilmaz}}, \bibinfo
  {author} {\bibfnamefont {Y.~Y.}\ \bibnamefont {Pai}}, \bibinfo {author}
  {\bibfnamefont {C.~E.}\ \bibnamefont {Marvinney}}, \bibinfo {author}
  {\bibfnamefont {A.}~\bibnamefont {Said}}, \bibinfo {author} {\bibfnamefont
  {Q.~W.}\ \bibnamefont {Yin}}, \bibinfo {author} {\bibfnamefont {C.~S.}\
  \bibnamefont {Gong}}, \bibinfo {author} {\bibfnamefont {Z.~J.}\ \bibnamefont
  {Tu}}, \bibinfo {author} {\bibfnamefont {E.}~\bibnamefont {Vescovo}},
  \bibinfo {author} {\bibfnamefont {C.~S.}\ \bibnamefont {Nelson}}, \bibinfo
  {author} {\bibfnamefont {R.~G.}\ \bibnamefont {Moore}}, \bibinfo {author}
  {\bibfnamefont {S.}~\bibnamefont {Murakami}}, \bibinfo {author}
  {\bibfnamefont {H.~C.}\ \bibnamefont {Lei}}, \bibinfo {author} {\bibfnamefont
  {H.~N.}\ \bibnamefont {Lee}}, \bibinfo {author} {\bibfnamefont {B.~J.}\
  \bibnamefont {Lawrie}}, \ and\ \bibinfo {author} {\bibfnamefont
  {H.}~\bibnamefont {Miao}},\ }\href {\doibase 10.1103/PhysRevX.11.031050}
  {\bibfield  {journal} {\bibinfo  {journal} {Phys. Rev. X}\ }\textbf {\bibinfo
  {volume} {11}},\ \bibinfo {pages} {031050} (\bibinfo {year}
  {2021})}\BibitemShut {NoStop}%
\bibitem [{\citenamefont {Moreo}\ \emph {et~al.}(1999)\citenamefont {Moreo},
  \citenamefont {Yunoki},\ and\ \citenamefont {Dagotto}}]{Moreo1999}%
  \BibitemOpen
  \bibfield  {author} {\bibinfo {author} {\bibfnamefont {A.}~\bibnamefont
  {Moreo}}, \bibinfo {author} {\bibfnamefont {S.}~\bibnamefont {Yunoki}}, \
  and\ \bibinfo {author} {\bibfnamefont {E.}~\bibnamefont {Dagotto}},\ }\href
  {\doibase 10.1126/science.283.5410.2034} {\bibfield  {journal} {\bibinfo
  {journal} {Science}\ }\textbf {\bibinfo {volume} {283}},\ \bibinfo {pages}
  {2034} (\bibinfo {year} {1999})}\BibitemShut {NoStop}%
\bibitem [{\citenamefont {Kuroda}\ \emph {et~al.}(2017)\citenamefont {Kuroda},
  \citenamefont {Tomita}, \citenamefont {Suzuki}, \citenamefont {Bareille},
  \citenamefont {Nugroho}, \citenamefont {Goswami}, \citenamefont {Ochi},
  \citenamefont {Ikhlas}, \citenamefont {Nakayama}, \citenamefont {Akebi},
  \citenamefont {Noguchi}, \citenamefont {Ishii}, \citenamefont {Inami},
  \citenamefont {Ono}, \citenamefont {Kumigashira}, \citenamefont {Varykhalov},
  \citenamefont {Muro}, \citenamefont {Koretsune}, \citenamefont {Arita},
  \citenamefont {Shin}, \citenamefont {Kondo},\ and\ \citenamefont
  {Nakatsuji}}]{Kuroda2017}%
  \BibitemOpen
  \bibfield  {author} {\bibinfo {author} {\bibfnamefont {K.}~\bibnamefont
  {Kuroda}}, \bibinfo {author} {\bibfnamefont {T.}~\bibnamefont {Tomita}},
  \bibinfo {author} {\bibfnamefont {M.-T.}\ \bibnamefont {Suzuki}}, \bibinfo
  {author} {\bibfnamefont {C.}~\bibnamefont {Bareille}}, \bibinfo {author}
  {\bibfnamefont {A.~Â.~A.}\ \bibnamefont {Nugroho}}, \bibinfo {author}
  {\bibfnamefont {P.}~\bibnamefont {Goswami}}, \bibinfo {author} {\bibfnamefont
  {M.}~\bibnamefont {Ochi}}, \bibinfo {author} {\bibfnamefont {M.}~\bibnamefont
  {Ikhlas}}, \bibinfo {author} {\bibfnamefont {M.}~\bibnamefont {Nakayama}},
  \bibinfo {author} {\bibfnamefont {S.}~\bibnamefont {Akebi}}, \bibinfo
  {author} {\bibfnamefont {R.}~\bibnamefont {Noguchi}}, \bibinfo {author}
  {\bibfnamefont {R.}~\bibnamefont {Ishii}}, \bibinfo {author} {\bibfnamefont
  {N.}~\bibnamefont {Inami}}, \bibinfo {author} {\bibfnamefont
  {K.}~\bibnamefont {Ono}}, \bibinfo {author} {\bibfnamefont {H.}~\bibnamefont
  {Kumigashira}}, \bibinfo {author} {\bibfnamefont {A.}~\bibnamefont
  {Varykhalov}}, \bibinfo {author} {\bibfnamefont {T.}~\bibnamefont {Muro}},
  \bibinfo {author} {\bibfnamefont {T.}~\bibnamefont {Koretsune}}, \bibinfo
  {author} {\bibfnamefont {R.}~\bibnamefont {Arita}}, \bibinfo {author}
  {\bibfnamefont {S.}~\bibnamefont {Shin}}, \bibinfo {author} {\bibfnamefont
  {T.}~\bibnamefont {Kondo}}, \ and\ \bibinfo {author} {\bibfnamefont
  {S.}~\bibnamefont {Nakatsuji}},\ }\href {\doibase 10.1038/nmat4987}
  {\bibfield  {journal} {\bibinfo  {journal} {Nat. Mater.}\ }\textbf {\bibinfo
  {volume} {16}},\ \bibinfo {pages} {1090} (\bibinfo {year}
  {2017})}\BibitemShut {NoStop}%
\bibitem [{\citenamefont {Chen}\ \emph {et~al.}(2021)\citenamefont {Chen},
  \citenamefont {Tomita}, \citenamefont {Minami}, \citenamefont {Fu},
  \citenamefont {Koretsune}, \citenamefont {Kitatani}, \citenamefont
  {Muhammad}, \citenamefont {Nishio-Hamane}, \citenamefont {Ishii},
  \citenamefont {Ishii} \emph {et~al.}}]{chen2021anomalous}%
  \BibitemOpen
  \bibfield  {author} {\bibinfo {author} {\bibfnamefont {T.}~\bibnamefont
  {Chen}}, \bibinfo {author} {\bibfnamefont {T.}~\bibnamefont {Tomita}},
  \bibinfo {author} {\bibfnamefont {S.}~\bibnamefont {Minami}}, \bibinfo
  {author} {\bibfnamefont {M.}~\bibnamefont {Fu}}, \bibinfo {author}
  {\bibfnamefont {T.}~\bibnamefont {Koretsune}}, \bibinfo {author}
  {\bibfnamefont {M.}~\bibnamefont {Kitatani}}, \bibinfo {author}
  {\bibfnamefont {I.}~\bibnamefont {Muhammad}}, \bibinfo {author}
  {\bibfnamefont {D.}~\bibnamefont {Nishio-Hamane}}, \bibinfo {author}
  {\bibfnamefont {R.}~\bibnamefont {Ishii}}, \bibinfo {author} {\bibfnamefont
  {F.}~\bibnamefont {Ishii}},  \emph {et~al.},\ }\href@noop {} {\bibfield
  {journal} {\bibinfo  {journal} {Nat. Commun.}\ }\textbf {\bibinfo {volume}
  {12}},\ \bibinfo {pages} {572} (\bibinfo {year} {2021})}\BibitemShut
  {NoStop}%
\bibitem [{\citenamefont {Cable}\ \emph {et~al.}(1993)\citenamefont {Cable},
  \citenamefont {Wakabayashi},\ and\ \citenamefont {Radhakrishna}}]{CABLE1993}%
  \BibitemOpen
  \bibfield  {author} {\bibinfo {author} {\bibfnamefont {J.}~\bibnamefont
  {Cable}}, \bibinfo {author} {\bibfnamefont {N.}~\bibnamefont {Wakabayashi}},
  \ and\ \bibinfo {author} {\bibfnamefont {P.}~\bibnamefont {Radhakrishna}},\
  }\href {\doibase https://doi.org/10.1016/0038-1098(93)90400-H} {\bibfield
  {journal} {\bibinfo  {journal} {Solid State Commun.}\ }\textbf {\bibinfo
  {volume} {88}},\ \bibinfo {pages} {161} (\bibinfo {year} {1993})}\BibitemShut
  {NoStop}%
\bibitem [{\citenamefont {Song}\ \emph {et~al.}(2020)\citenamefont {Song},
  \citenamefont {Hao}, \citenamefont {Wang}, \citenamefont {Zhang},
  \citenamefont {Huang}, \citenamefont {Xing},\ and\ \citenamefont
  {Chen}}]{Song2020}%
  \BibitemOpen
  \bibfield  {author} {\bibinfo {author} {\bibfnamefont {Y.}~\bibnamefont
  {Song}}, \bibinfo {author} {\bibfnamefont {Y.}~\bibnamefont {Hao}}, \bibinfo
  {author} {\bibfnamefont {S.}~\bibnamefont {Wang}}, \bibinfo {author}
  {\bibfnamefont {J.}~\bibnamefont {Zhang}}, \bibinfo {author} {\bibfnamefont
  {Q.}~\bibnamefont {Huang}}, \bibinfo {author} {\bibfnamefont
  {X.}~\bibnamefont {Xing}}, \ and\ \bibinfo {author} {\bibfnamefont
  {J.}~\bibnamefont {Chen}},\ }\href {\doibase 10.1103/PhysRevB.101.144422}
  {\bibfield  {journal} {\bibinfo  {journal} {Phys. Rev. B}\ }\textbf {\bibinfo
  {volume} {101}},\ \bibinfo {pages} {144422} (\bibinfo {year}
  {2020})}\BibitemShut {NoStop}%
\bibitem [{\citenamefont {Tomiyoshi}\ and\ \citenamefont
  {Yamaguchi}(1982)}]{tomiyoshi1982}%
  \BibitemOpen
  \bibfield  {author} {\bibinfo {author} {\bibfnamefont {S.}~\bibnamefont
  {Tomiyoshi}}\ and\ \bibinfo {author} {\bibfnamefont {Y.}~\bibnamefont
  {Yamaguchi}},\ }\href {\doibase https://doi.org/10.1143/JPSJ.51.2478}
  {\bibfield  {journal} {\bibinfo  {journal} {J. Phys. Soc. Japan}\ }\textbf
  {\bibinfo {volume} {51}},\ \bibinfo {pages} {2478} (\bibinfo {year}
  {1982})}\BibitemShut {NoStop}%
\bibitem [{\citenamefont {Tomiyoshi}(1982)}]{Tomiyoshi1982v2}%
  \BibitemOpen
  \bibfield  {author} {\bibinfo {author} {\bibfnamefont {S.}~\bibnamefont
  {Tomiyoshi}},\ }\href {\doibase 10.1143/JPSJ.51.803} {\bibfield  {journal}
  {\bibinfo  {journal} {J. Phys. Soc. Japan}\ }\textbf {\bibinfo {volume}
  {51}},\ \bibinfo {pages} {803} (\bibinfo {year} {1982})}\BibitemShut
  {NoStop}%
\bibitem [{\citenamefont {Ohmori}\ \emph {et~al.}(1987)\citenamefont {Ohmori},
  \citenamefont {Tomiyoshi}, \citenamefont {Yamauchi},\ and\ \citenamefont
  {Yamamoto}}]{OHMORI1987}%
  \BibitemOpen
  \bibfield  {author} {\bibinfo {author} {\bibfnamefont {H.}~\bibnamefont
  {Ohmori}}, \bibinfo {author} {\bibfnamefont {S.}~\bibnamefont {Tomiyoshi}},
  \bibinfo {author} {\bibfnamefont {H.}~\bibnamefont {Yamauchi}}, \ and\
  \bibinfo {author} {\bibfnamefont {H.}~\bibnamefont {Yamamoto}},\ }\href
  {\doibase https://doi.org/10.1016/0304-8853(87)90427-6} {\bibfield  {journal}
  {\bibinfo  {journal} {J. Magn. Magn. Mater.}\ }\textbf {\bibinfo {volume}
  {70}},\ \bibinfo {pages} {249} (\bibinfo {year} {1987})}\BibitemShut
  {NoStop}%
\bibitem [{\citenamefont {Brown}\ \emph
  {et~al.}(1990{\natexlab{a}})\citenamefont {Brown}, \citenamefont {Nunez},
  \citenamefont {Tasset}, \citenamefont {Forsyth},\ and\ \citenamefont
  {Radhakrishna}}]{Brown1990}%
  \BibitemOpen
  \bibfield  {author} {\bibinfo {author} {\bibfnamefont {P.~J.}\ \bibnamefont
  {Brown}}, \bibinfo {author} {\bibfnamefont {V.}~\bibnamefont {Nunez}},
  \bibinfo {author} {\bibfnamefont {F.}~\bibnamefont {Tasset}}, \bibinfo
  {author} {\bibfnamefont {J.~B.}\ \bibnamefont {Forsyth}}, \ and\ \bibinfo
  {author} {\bibfnamefont {P.}~\bibnamefont {Radhakrishna}},\ }\href {\doibase
  10.1088/0953-8984/2/47/015} {\bibfield  {journal} {\bibinfo  {journal} {J.
  Condens. Matter Phys.}\ }\textbf {\bibinfo {volume} {2}},\ \bibinfo {pages}
  {9409} (\bibinfo {year} {1990}{\natexlab{a}})}\BibitemShut {NoStop}%
\bibitem [{\citenamefont {Ikhlas}\ \emph {et~al.}(2017)\citenamefont {Ikhlas},
  \citenamefont {Tomita}, \citenamefont {Koretsune}, \citenamefont {Suzuki},
  \citenamefont {Nishio-Hamane}, \citenamefont {Arita}, \citenamefont {Otani},\
  and\ \citenamefont {Nakatsuji}}]{Ikhlas2017}%
  \BibitemOpen
  \bibfield  {author} {\bibinfo {author} {\bibfnamefont {M.}~\bibnamefont
  {Ikhlas}}, \bibinfo {author} {\bibfnamefont {T.}~\bibnamefont {Tomita}},
  \bibinfo {author} {\bibfnamefont {T.}~\bibnamefont {Koretsune}}, \bibinfo
  {author} {\bibfnamefont {M.-T.}\ \bibnamefont {Suzuki}}, \bibinfo {author}
  {\bibfnamefont {D.}~\bibnamefont {Nishio-Hamane}}, \bibinfo {author}
  {\bibfnamefont {R.}~\bibnamefont {Arita}}, \bibinfo {author} {\bibfnamefont
  {Y.}~\bibnamefont {Otani}}, \ and\ \bibinfo {author} {\bibfnamefont
  {S.}~\bibnamefont {Nakatsuji}},\ }\href {\doibase 10.1038/nphys4181}
  {\bibfield  {journal} {\bibinfo  {journal} {Nat. Phys.}\ }\textbf {\bibinfo
  {volume} {13}},\ \bibinfo {pages} {1085} (\bibinfo {year}
  {2017})}\BibitemShut {NoStop}%
\bibitem [{\citenamefont {Kimata}\ \emph {et~al.}(2019)\citenamefont {Kimata},
  \citenamefont {Chen}, \citenamefont {Kondou}, \citenamefont {Sugimoto},
  \citenamefont {Muduli}, \citenamefont {Ikhlas}, \citenamefont {Omori},
  \citenamefont {Tomita}, \citenamefont {MacDonald}, \citenamefont
  {Nakatsuji},\ and\ \citenamefont {Otani}}]{Kimata2019}%
  \BibitemOpen
  \bibfield  {author} {\bibinfo {author} {\bibfnamefont {M.}~\bibnamefont
  {Kimata}}, \bibinfo {author} {\bibfnamefont {H.}~\bibnamefont {Chen}},
  \bibinfo {author} {\bibfnamefont {K.}~\bibnamefont {Kondou}}, \bibinfo
  {author} {\bibfnamefont {S.}~\bibnamefont {Sugimoto}}, \bibinfo {author}
  {\bibfnamefont {P.~K.}\ \bibnamefont {Muduli}}, \bibinfo {author}
  {\bibfnamefont {M.}~\bibnamefont {Ikhlas}}, \bibinfo {author} {\bibfnamefont
  {Y.}~\bibnamefont {Omori}}, \bibinfo {author} {\bibfnamefont
  {T.}~\bibnamefont {Tomita}}, \bibinfo {author} {\bibfnamefont {A.~H.}\
  \bibnamefont {MacDonald}}, \bibinfo {author} {\bibfnamefont {S.}~\bibnamefont
  {Nakatsuji}}, \ and\ \bibinfo {author} {\bibfnamefont {Y.}~\bibnamefont
  {Otani}},\ }\href {\doibase 10.1038/s41586-018-0853-0} {\bibfield  {journal}
  {\bibinfo  {journal} {Nature}\ }\textbf {\bibinfo {volume} {565}},\ \bibinfo
  {pages} {627} (\bibinfo {year} {2019})}\BibitemShut {NoStop}%
\bibitem [{\citenamefont {Li}\ \emph {et~al.}(2018)\citenamefont {Li},
  \citenamefont {Zhuang}, \citenamefont {Wang}, \citenamefont {Feng},
  \citenamefont {Gao}, \citenamefont {Xu}, \citenamefont {Hao}, \citenamefont
  {Wang}, \citenamefont {Zhang}, \citenamefont {Wu}, \citenamefont {Dou},
  \citenamefont {Chen}, \citenamefont {Hu},\ and\ \citenamefont {Du}}]{Li2018}%
  \BibitemOpen
  \bibfield  {author} {\bibinfo {author} {\bibfnamefont {Z.}~\bibnamefont
  {Li}}, \bibinfo {author} {\bibfnamefont {J.}~\bibnamefont {Zhuang}}, \bibinfo
  {author} {\bibfnamefont {L.}~\bibnamefont {Wang}}, \bibinfo {author}
  {\bibfnamefont {H.}~\bibnamefont {Feng}}, \bibinfo {author} {\bibfnamefont
  {Q.}~\bibnamefont {Gao}}, \bibinfo {author} {\bibfnamefont {X.}~\bibnamefont
  {Xu}}, \bibinfo {author} {\bibfnamefont {W.}~\bibnamefont {Hao}}, \bibinfo
  {author} {\bibfnamefont {X.}~\bibnamefont {Wang}}, \bibinfo {author}
  {\bibfnamefont {C.}~\bibnamefont {Zhang}}, \bibinfo {author} {\bibfnamefont
  {K.}~\bibnamefont {Wu}}, \bibinfo {author} {\bibfnamefont {S.~X.}\
  \bibnamefont {Dou}}, \bibinfo {author} {\bibfnamefont {L.}~\bibnamefont
  {Chen}}, \bibinfo {author} {\bibfnamefont {Z.}~\bibnamefont {Hu}}, \ and\
  \bibinfo {author} {\bibfnamefont {Y.}~\bibnamefont {Du}},\ }\href {\doibase
  10.1126/sciadv.aau4511} {\bibfield  {journal} {\bibinfo  {journal} {Sci.
  Adv.}\ }\textbf {\bibinfo {volume} {4}},\ \bibinfo {pages} {eaau4511}
  (\bibinfo {year} {2018})}\BibitemShut {NoStop}%
\bibitem [{\citenamefont {Ghimire}\ and\ \citenamefont
  {Mazin}(2020)}]{ghimire2020}%
  \BibitemOpen
  \bibfield  {author} {\bibinfo {author} {\bibfnamefont {N.~J.}\ \bibnamefont
  {Ghimire}}\ and\ \bibinfo {author} {\bibfnamefont {I.~I.}\ \bibnamefont
  {Mazin}},\ }\href {\doibase https://doi.org/10.1038/s41563-019-0589-8}
  {\bibfield  {journal} {\bibinfo  {journal} {Nat. Mater.}\ }\textbf {\bibinfo
  {volume} {19}},\ \bibinfo {pages} {137} (\bibinfo {year} {2020})}\BibitemShut
  {NoStop}%
\bibitem [{\citenamefont {Fawcett}(1988)}]{RevModPhys.60.209}%
  \BibitemOpen
  \bibfield  {author} {\bibinfo {author} {\bibfnamefont {E.}~\bibnamefont
  {Fawcett}},\ }\href {\doibase 10.1103/RevModPhys.60.209} {\bibfield
  {journal} {\bibinfo  {journal} {Rev. Mod. Phys.}\ }\textbf {\bibinfo {volume}
  {60}},\ \bibinfo {pages} {209} (\bibinfo {year} {1988})}\BibitemShut
  {NoStop}%
\bibitem [{\citenamefont {Ortiz}\ \emph
  {et~al.}(2020{\natexlab{b}})\citenamefont {Ortiz}, \citenamefont {Teicher},
  \citenamefont {Hu}, \citenamefont {Zuo}, \citenamefont {Sarte}, \citenamefont
  {Schueller}, \citenamefont {Abeykoon}, \citenamefont {Krogstad},
  \citenamefont {Rosenkranz}, \citenamefont {Osborn}, \citenamefont {Seshadri},
  \citenamefont {Balents}, \citenamefont {He},\ and\ \citenamefont
  {Wilson}}]{Ortiz2020}%
  \BibitemOpen
  \bibfield  {author} {\bibinfo {author} {\bibfnamefont {B.~R.}\ \bibnamefont
  {Ortiz}}, \bibinfo {author} {\bibfnamefont {S.~M.~L.}\ \bibnamefont
  {Teicher}}, \bibinfo {author} {\bibfnamefont {Y.}~\bibnamefont {Hu}},
  \bibinfo {author} {\bibfnamefont {J.~L.}\ \bibnamefont {Zuo}}, \bibinfo
  {author} {\bibfnamefont {P.~M.}\ \bibnamefont {Sarte}}, \bibinfo {author}
  {\bibfnamefont {E.~C.}\ \bibnamefont {Schueller}}, \bibinfo {author}
  {\bibfnamefont {A.~M.~M.}\ \bibnamefont {Abeykoon}}, \bibinfo {author}
  {\bibfnamefont {M.~J.}\ \bibnamefont {Krogstad}}, \bibinfo {author}
  {\bibfnamefont {S.}~\bibnamefont {Rosenkranz}}, \bibinfo {author}
  {\bibfnamefont {R.}~\bibnamefont {Osborn}}, \bibinfo {author} {\bibfnamefont
  {R.}~\bibnamefont {Seshadri}}, \bibinfo {author} {\bibfnamefont
  {L.}~\bibnamefont {Balents}}, \bibinfo {author} {\bibfnamefont
  {J.}~\bibnamefont {He}}, \ and\ \bibinfo {author} {\bibfnamefont {S.~D.}\
  \bibnamefont {Wilson}},\ }\href {\doibase 10.1103/PhysRevLett.125.247002}
  {\bibfield  {journal} {\bibinfo  {journal} {Phys. Rev. Lett.}\ }\textbf
  {\bibinfo {volume} {125}},\ \bibinfo {pages} {247002} (\bibinfo {year}
  {2020}{\natexlab{b}})}\BibitemShut {NoStop}%
\bibitem [{\citenamefont {Teng}\ \emph {et~al.}(2022)\citenamefont {Teng},
  \citenamefont {Chen}, \citenamefont {Ye}, \citenamefont {Rosenberg},
  \citenamefont {Liu}, \citenamefont {Yin}, \citenamefont {Jiang},
  \citenamefont {Oh}, \citenamefont {Hasan}, \citenamefont {Kelly~J},
  \citenamefont {Gao}, \citenamefont {Xie}, \citenamefont {Hashimoto},
  \citenamefont {Lu}, \citenamefont {Jozwiak}, \citenamefont {Bostwick},
  \citenamefont {Rotenberg}, \citenamefont {Birgeneau}, \citenamefont {Chu},
  \citenamefont {Yi},\ and\ \citenamefont {Dai}}]{teng2022}%
  \BibitemOpen
  \bibfield  {author} {\bibinfo {author} {\bibfnamefont {X.}~\bibnamefont
  {Teng}}, \bibinfo {author} {\bibfnamefont {L.}~\bibnamefont {Chen}}, \bibinfo
  {author} {\bibfnamefont {F.}~\bibnamefont {Ye}}, \bibinfo {author}
  {\bibfnamefont {E.}~\bibnamefont {Rosenberg}}, \bibinfo {author}
  {\bibfnamefont {Z.}~\bibnamefont {Liu}}, \bibinfo {author} {\bibfnamefont
  {J.-X.}\ \bibnamefont {Yin}}, \bibinfo {author} {\bibfnamefont {Y.-X.}\
  \bibnamefont {Jiang}}, \bibinfo {author} {\bibfnamefont {J.~S.}\ \bibnamefont
  {Oh}}, \bibinfo {author} {\bibfnamefont {M.~Z.}\ \bibnamefont {Hasan}},
  \bibinfo {author} {\bibfnamefont {N.}~\bibnamefont {Kelly~J}}, \bibinfo
  {author} {\bibfnamefont {B.}~\bibnamefont {Gao}}, \bibinfo {author}
  {\bibfnamefont {Y.}~\bibnamefont {Xie}}, \bibinfo {author} {\bibfnamefont
  {M.}~\bibnamefont {Hashimoto}}, \bibinfo {author} {\bibfnamefont
  {D.}~\bibnamefont {Lu}}, \bibinfo {author} {\bibfnamefont {C.}~\bibnamefont
  {Jozwiak}}, \bibinfo {author} {\bibfnamefont {A.}~\bibnamefont {Bostwick}},
  \bibinfo {author} {\bibfnamefont {E.}~\bibnamefont {Rotenberg}}, \bibinfo
  {author} {\bibfnamefont {R.~J.}\ \bibnamefont {Birgeneau}}, \bibinfo {author}
  {\bibfnamefont {J.-H.}\ \bibnamefont {Chu}}, \bibinfo {author} {\bibfnamefont
  {M.}~\bibnamefont {Yi}}, \ and\ \bibinfo {author} {\bibfnamefont
  {P.}~\bibnamefont {Dai}},\ }\href {\doibase
  https://doi.org/10.1038/s41586-022-05034-z} {\bibfield  {journal} {\bibinfo
  {journal} {Nature}\ }\textbf {\bibinfo {volume} {609}},\ \bibinfo {pages}
  {490} (\bibinfo {year} {2022})}\BibitemShut {NoStop}%
\bibitem [{\citenamefont {Teng}\ \emph {et~al.}(2023)\citenamefont {Teng},
  \citenamefont {Oh}, \citenamefont {Tan}, \citenamefont {Chen}, \citenamefont
  {Huang}, \citenamefont {Gao}, \citenamefont {Yin}, \citenamefont {Chu},
  \citenamefont {Hashimoto}, \citenamefont {Lu}, \citenamefont {Jozwiak},
  \citenamefont {Bostwick}, \citenamefont {Rotenberg}, \citenamefont
  {Granroth}, \citenamefont {Yan}, \citenamefont {Birgeneau}, \citenamefont
  {Dai},\ and\ \citenamefont {Yi}}]{teng2023}%
  \BibitemOpen
  \bibfield  {author} {\bibinfo {author} {\bibfnamefont {X.}~\bibnamefont
  {Teng}}, \bibinfo {author} {\bibfnamefont {J.~S.}\ \bibnamefont {Oh}},
  \bibinfo {author} {\bibfnamefont {H.}~\bibnamefont {Tan}}, \bibinfo {author}
  {\bibfnamefont {L.}~\bibnamefont {Chen}}, \bibinfo {author} {\bibfnamefont
  {J.}~\bibnamefont {Huang}}, \bibinfo {author} {\bibfnamefont
  {B.}~\bibnamefont {Gao}}, \bibinfo {author} {\bibfnamefont {J.-X.}\
  \bibnamefont {Yin}}, \bibinfo {author} {\bibfnamefont {J.-H.}\ \bibnamefont
  {Chu}}, \bibinfo {author} {\bibfnamefont {M.}~\bibnamefont {Hashimoto}},
  \bibinfo {author} {\bibfnamefont {D.}~\bibnamefont {Lu}}, \bibinfo {author}
  {\bibfnamefont {C.}~\bibnamefont {Jozwiak}}, \bibinfo {author} {\bibfnamefont
  {A.}~\bibnamefont {Bostwick}}, \bibinfo {author} {\bibfnamefont
  {E.}~\bibnamefont {Rotenberg}}, \bibinfo {author} {\bibfnamefont {G.~E.}\
  \bibnamefont {Granroth}}, \bibinfo {author} {\bibfnamefont {B.}~\bibnamefont
  {Yan}}, \bibinfo {author} {\bibfnamefont {R.~J.}\ \bibnamefont {Birgeneau}},
  \bibinfo {author} {\bibfnamefont {P.}~\bibnamefont {Dai}}, \ and\ \bibinfo
  {author} {\bibfnamefont {M.}~\bibnamefont {Yi}},\ }\href {\doibase
  https://doi.org/10.1038/s41567-023-01985-w} {\bibfield  {journal} {\bibinfo
  {journal} {Nat. Phys.}\ }\textbf {\bibinfo {volume} {19}},\ \bibinfo {pages}
  {814} (\bibinfo {year} {2023})}\BibitemShut {NoStop}%
\bibitem [{\citenamefont {Chen}\ \emph {et~al.}(2020)\citenamefont {Chen},
  \citenamefont {Gaudet}, \citenamefont {Dasgupta}, \citenamefont {Marcus},
  \citenamefont {Lin}, \citenamefont {Chen}, \citenamefont {Tomita},
  \citenamefont {Ikhlas}, \citenamefont {Zhao}, \citenamefont {Chen},
  \citenamefont {Stone}, \citenamefont {Tchernyshyov}, \citenamefont
  {Nakatsuji},\ and\ \citenamefont {Broholm}}]{PhysRevB.102.054403}%
  \BibitemOpen
  \bibfield  {author} {\bibinfo {author} {\bibfnamefont {Y.}~\bibnamefont
  {Chen}}, \bibinfo {author} {\bibfnamefont {J.}~\bibnamefont {Gaudet}},
  \bibinfo {author} {\bibfnamefont {S.}~\bibnamefont {Dasgupta}}, \bibinfo
  {author} {\bibfnamefont {G.~G.}\ \bibnamefont {Marcus}}, \bibinfo {author}
  {\bibfnamefont {J.}~\bibnamefont {Lin}}, \bibinfo {author} {\bibfnamefont
  {T.}~\bibnamefont {Chen}}, \bibinfo {author} {\bibfnamefont {T.}~\bibnamefont
  {Tomita}}, \bibinfo {author} {\bibfnamefont {M.}~\bibnamefont {Ikhlas}},
  \bibinfo {author} {\bibfnamefont {Y.}~\bibnamefont {Zhao}}, \bibinfo {author}
  {\bibfnamefont {W.~C.}\ \bibnamefont {Chen}}, \bibinfo {author}
  {\bibfnamefont {M.~B.}\ \bibnamefont {Stone}}, \bibinfo {author}
  {\bibfnamefont {O.}~\bibnamefont {Tchernyshyov}}, \bibinfo {author}
  {\bibfnamefont {S.}~\bibnamefont {Nakatsuji}}, \ and\ \bibinfo {author}
  {\bibfnamefont {C.}~\bibnamefont {Broholm}},\ }\href {\doibase
  10.1103/PhysRevB.102.054403} {\bibfield  {journal} {\bibinfo  {journal}
  {Phys. Rev. B}\ }\textbf {\bibinfo {volume} {102}},\ \bibinfo {pages}
  {054403} (\bibinfo {year} {2020})}\BibitemShut {NoStop}%
\bibitem [{\citenamefont {Brown}\ \emph
  {et~al.}(1990{\natexlab{b}})\citenamefont {Brown}, \citenamefont {Nunez},
  \citenamefont {Tasset}, \citenamefont {Forsyth},\ and\ \citenamefont
  {Radhakrishna}}]{Brown_1990}%
  \BibitemOpen
  \bibfield  {author} {\bibinfo {author} {\bibfnamefont {P.~J.}\ \bibnamefont
  {Brown}}, \bibinfo {author} {\bibfnamefont {V.}~\bibnamefont {Nunez}},
  \bibinfo {author} {\bibfnamefont {F.}~\bibnamefont {Tasset}}, \bibinfo
  {author} {\bibfnamefont {J.~B.}\ \bibnamefont {Forsyth}}, \ and\ \bibinfo
  {author} {\bibfnamefont {P.}~\bibnamefont {Radhakrishna}},\ }\href {\doibase
  10.1088/0953-8984/2/47/015} {\bibfield  {journal} {\bibinfo  {journal} {J.
  Condens. Matter Phys.}\ }\textbf {\bibinfo {volume} {2}},\ \bibinfo {pages}
  {9409} (\bibinfo {year} {1990}{\natexlab{b}})}\BibitemShut {NoStop}%
\bibitem [{\citenamefont {Soh}\ \emph {et~al.}(2020)\citenamefont {Soh},
  \citenamefont {de~Juan}, \citenamefont {Qureshi}, \citenamefont {Jacobsen},
  \citenamefont {Wang}, \citenamefont {Guo},\ and\ \citenamefont
  {Boothroyd}}]{PhysRevB.101.140411}%
  \BibitemOpen
  \bibfield  {author} {\bibinfo {author} {\bibfnamefont {J.-R.}\ \bibnamefont
  {Soh}}, \bibinfo {author} {\bibfnamefont {F.}~\bibnamefont {de~Juan}},
  \bibinfo {author} {\bibfnamefont {N.}~\bibnamefont {Qureshi}}, \bibinfo
  {author} {\bibfnamefont {H.}~\bibnamefont {Jacobsen}}, \bibinfo {author}
  {\bibfnamefont {H.-Y.}\ \bibnamefont {Wang}}, \bibinfo {author}
  {\bibfnamefont {Y.-F.}\ \bibnamefont {Guo}}, \ and\ \bibinfo {author}
  {\bibfnamefont {A.~T.}\ \bibnamefont {Boothroyd}},\ }\href {\doibase
  10.1103/PhysRevB.101.140411} {\bibfield  {journal} {\bibinfo  {journal}
  {Phys. Rev. B}\ }\textbf {\bibinfo {volume} {101}},\ \bibinfo {pages}
  {140411} (\bibinfo {year} {2020})}\BibitemShut {NoStop}%
\bibitem [{\citenamefont {Khadka}\ \emph {et~al.}(2020)\citenamefont {Khadka},
  \citenamefont {Thapaliya}, \citenamefont {Hurtado~Parra}, \citenamefont
  {Han}, \citenamefont {Wen}, \citenamefont {Need}, \citenamefont {Khanal},
  \citenamefont {Wang}, \citenamefont {Zang}, \citenamefont {Kikkawa},
  \citenamefont {Wu},\ and\ \citenamefont {Huang}}]{Khadkaeabc1977}%
  \BibitemOpen
  \bibfield  {author} {\bibinfo {author} {\bibfnamefont {D.}~\bibnamefont
  {Khadka}}, \bibinfo {author} {\bibfnamefont {T.~R.}\ \bibnamefont
  {Thapaliya}}, \bibinfo {author} {\bibfnamefont {S.}~\bibnamefont
  {Hurtado~Parra}}, \bibinfo {author} {\bibfnamefont {X.}~\bibnamefont {Han}},
  \bibinfo {author} {\bibfnamefont {J.}~\bibnamefont {Wen}}, \bibinfo {author}
  {\bibfnamefont {R.~F.}\ \bibnamefont {Need}}, \bibinfo {author}
  {\bibfnamefont {P.}~\bibnamefont {Khanal}}, \bibinfo {author} {\bibfnamefont
  {W.}~\bibnamefont {Wang}}, \bibinfo {author} {\bibfnamefont {J.}~\bibnamefont
  {Zang}}, \bibinfo {author} {\bibfnamefont {J.~M.}\ \bibnamefont {Kikkawa}},
  \bibinfo {author} {\bibfnamefont {L.}~\bibnamefont {Wu}}, \ and\ \bibinfo
  {author} {\bibfnamefont {S.~X.}\ \bibnamefont {Huang}},\ }\href
  {https://advances.sciencemag.org/content/6/35/eabc1977} {\bibfield  {journal}
  {\bibinfo  {journal} {Sci. Adv.}\ }\textbf {\bibinfo {volume} {6}} (\bibinfo
  {year} {2020})}\BibitemShut {NoStop}%
\bibitem [{\citenamefont {Dasgupta}\ and\ \citenamefont
  {Tchernyshyov}(2020)}]{PhysRevB.102.144417}%
  \BibitemOpen
  \bibfield  {author} {\bibinfo {author} {\bibfnamefont {S.}~\bibnamefont
  {Dasgupta}}\ and\ \bibinfo {author} {\bibfnamefont {O.}~\bibnamefont
  {Tchernyshyov}},\ }\href {\doibase 10.1103/PhysRevB.102.144417} {\bibfield
  {journal} {\bibinfo  {journal} {Phys. Rev. B}\ }\textbf {\bibinfo {volume}
  {102}},\ \bibinfo {pages} {144417} (\bibinfo {year} {2020})}\BibitemShut
  {NoStop}%
\bibitem [{\citenamefont {Ikhlas}\ \emph {et~al.}()\citenamefont {Ikhlas},
  \citenamefont {Tomita},\ and\ \citenamefont
  {Nakatsuji}}]{doi:10.7566/JPSCP.30.011177}%
  \BibitemOpen
  \bibfield  {author} {\bibinfo {author} {\bibfnamefont {M.}~\bibnamefont
  {Ikhlas}}, \bibinfo {author} {\bibfnamefont {T.}~\bibnamefont {Tomita}}, \
  and\ \bibinfo {author} {\bibfnamefont {S.}~\bibnamefont {Nakatsuji}},\
  }\enquote {\bibinfo {title} {Sample quality dependence of the magnetic
  properties in non-collinear antiferromagnet $\mathrm{Mn}_{3}\mathrm{Sn}$},}\
  in\ \href {https://journals.jps.jp/doi/abs/10.7566/JPSCP.30.011177} {\emph
  {\bibinfo {booktitle} {Proceedings of the International Conference on
  Strongly Correlated Electron Systems (SCES2019)}}}\BibitemShut {NoStop}%
\bibitem [{\citenamefont {Feng}\ \emph {et~al.}(2006)\citenamefont {Feng},
  \citenamefont {Li}, \citenamefont {Ren}, \citenamefont {Li}, \citenamefont
  {Li}, \citenamefont {Li}, \citenamefont {Zhang},\ and\ \citenamefont
  {Zhang}}]{PhysRevB.73.205105}%
  \BibitemOpen
  \bibfield  {author} {\bibinfo {author} {\bibfnamefont {W.~J.}\ \bibnamefont
  {Feng}}, \bibinfo {author} {\bibfnamefont {D.}~\bibnamefont {Li}}, \bibinfo
  {author} {\bibfnamefont {W.~J.}\ \bibnamefont {Ren}}, \bibinfo {author}
  {\bibfnamefont {Y.~B.}\ \bibnamefont {Li}}, \bibinfo {author} {\bibfnamefont
  {W.~F.}\ \bibnamefont {Li}}, \bibinfo {author} {\bibfnamefont
  {J.}~\bibnamefont {Li}}, \bibinfo {author} {\bibfnamefont {Y.~Q.}\
  \bibnamefont {Zhang}}, \ and\ \bibinfo {author} {\bibfnamefont {Z.~D.}\
  \bibnamefont {Zhang}},\ }\href {\doibase 10.1103/PhysRevB.73.205105}
  {\bibfield  {journal} {\bibinfo  {journal} {Phys. Rev. B}\ }\textbf {\bibinfo
  {volume} {73}},\ \bibinfo {pages} {205105} (\bibinfo {year}
  {2006})}\BibitemShut {NoStop}%
\bibitem [{\citenamefont {Li}\ \emph {et~al.}(2017)\citenamefont {Li},
  \citenamefont {Xu}, \citenamefont {Ding}, \citenamefont {Wang}, \citenamefont
  {Shen}, \citenamefont {Lu}, \citenamefont {Zhu},\ and\ \citenamefont
  {Behnia}}]{PhysRevLett.119.056601}%
  \BibitemOpen
  \bibfield  {author} {\bibinfo {author} {\bibfnamefont {X.}~\bibnamefont
  {Li}}, \bibinfo {author} {\bibfnamefont {L.}~\bibnamefont {Xu}}, \bibinfo
  {author} {\bibfnamefont {L.}~\bibnamefont {Ding}}, \bibinfo {author}
  {\bibfnamefont {J.}~\bibnamefont {Wang}}, \bibinfo {author} {\bibfnamefont
  {M.}~\bibnamefont {Shen}}, \bibinfo {author} {\bibfnamefont {X.}~\bibnamefont
  {Lu}}, \bibinfo {author} {\bibfnamefont {Z.}~\bibnamefont {Zhu}}, \ and\
  \bibinfo {author} {\bibfnamefont {K.}~\bibnamefont {Behnia}},\ }\href
  {\doibase 10.1103/PhysRevLett.119.056601} {\bibfield  {journal} {\bibinfo
  {journal} {Phys. Rev. Lett.}\ }\textbf {\bibinfo {volume} {119}},\ \bibinfo
  {pages} {056601} (\bibinfo {year} {2017})}\BibitemShut {NoStop}%
\bibitem [{\citenamefont {Sung}\ \emph {et~al.}(2018)\citenamefont {Sung},
  \citenamefont {Ronning}, \citenamefont {Thompson},\ and\ \citenamefont
  {Bauer}}]{doi:10.1063/1.5021133}%
  \BibitemOpen
  \bibfield  {author} {\bibinfo {author} {\bibfnamefont {N.~H.}\ \bibnamefont
  {Sung}}, \bibinfo {author} {\bibfnamefont {F.}~\bibnamefont {Ronning}},
  \bibinfo {author} {\bibfnamefont {J.~D.}\ \bibnamefont {Thompson}}, \ and\
  \bibinfo {author} {\bibfnamefont {E.~D.}\ \bibnamefont {Bauer}},\ }\href
  {\doibase 10.1063/1.5021133} {\bibfield  {journal} {\bibinfo  {journal}
  {Appl. Phys. Lett.}\ }\textbf {\bibinfo {volume} {112}},\ \bibinfo {pages}
  {132406} (\bibinfo {year} {2018})}\BibitemShut {NoStop}%
\bibitem [{\citenamefont {Ghimire}\ \emph {et~al.}(2020)\citenamefont
  {Ghimire}, \citenamefont {Dally}, \citenamefont {Poudel}, \citenamefont
  {Jones}, \citenamefont {Michel}, \citenamefont {Magar}, \citenamefont
  {Bleuel}, \citenamefont {McGuire}, \citenamefont {Jiang}, \citenamefont
  {Mitchell}, \citenamefont {Lynn},\ and\ \citenamefont
  {Mazin}}]{doi:10.1126/sciadv.abe2680}%
  \BibitemOpen
  \bibfield  {author} {\bibinfo {author} {\bibfnamefont {N.~J.}\ \bibnamefont
  {Ghimire}}, \bibinfo {author} {\bibfnamefont {R.~L.}\ \bibnamefont {Dally}},
  \bibinfo {author} {\bibfnamefont {L.}~\bibnamefont {Poudel}}, \bibinfo
  {author} {\bibfnamefont {D.~C.}\ \bibnamefont {Jones}}, \bibinfo {author}
  {\bibfnamefont {D.}~\bibnamefont {Michel}}, \bibinfo {author} {\bibfnamefont
  {N.~T.}\ \bibnamefont {Magar}}, \bibinfo {author} {\bibfnamefont
  {M.}~\bibnamefont {Bleuel}}, \bibinfo {author} {\bibfnamefont {M.~A.}\
  \bibnamefont {McGuire}}, \bibinfo {author} {\bibfnamefont {J.~S.}\
  \bibnamefont {Jiang}}, \bibinfo {author} {\bibfnamefont {J.~F.}\ \bibnamefont
  {Mitchell}}, \bibinfo {author} {\bibfnamefont {J.~W.}\ \bibnamefont {Lynn}},
  \ and\ \bibinfo {author} {\bibfnamefont {I.~I.}\ \bibnamefont {Mazin}},\
  }\href {\doibase 10.1126/sciadv.abe2680} {\bibfield  {journal} {\bibinfo
  {journal} {Sci. Adv.}\ }\textbf {\bibinfo {volume} {6}},\ \bibinfo {pages}
  {eabe2680} (\bibinfo {year} {2020})}\BibitemShut {NoStop}%
\bibitem [{\citenamefont {Park}\ \emph {et~al.}(2018)\citenamefont {Park},
  \citenamefont {Oh}, \citenamefont {Uhl{\'i}{\v{r}}ov{\'a}}, \citenamefont
  {Jackson}, \citenamefont {De{\'a}k}, \citenamefont {Szunyogh}, \citenamefont
  {Lee}, \citenamefont {Cho}, \citenamefont {Kim}, \citenamefont {Walker},
  \citenamefont {Adroja}, \citenamefont {Sechovsk{\'y}},\ and\ \citenamefont
  {Park}}]{Park2018}%
  \BibitemOpen
  \bibfield  {author} {\bibinfo {author} {\bibfnamefont {P.}~\bibnamefont
  {Park}}, \bibinfo {author} {\bibfnamefont {J.}~\bibnamefont {Oh}}, \bibinfo
  {author} {\bibfnamefont {K.}~\bibnamefont {Uhl{\'i}{\v{r}}ov{\'a}}}, \bibinfo
  {author} {\bibfnamefont {J.}~\bibnamefont {Jackson}}, \bibinfo {author}
  {\bibfnamefont {A.}~\bibnamefont {De{\'a}k}}, \bibinfo {author}
  {\bibfnamefont {L.}~\bibnamefont {Szunyogh}}, \bibinfo {author}
  {\bibfnamefont {K.~H.}\ \bibnamefont {Lee}}, \bibinfo {author} {\bibfnamefont
  {H.}~\bibnamefont {Cho}}, \bibinfo {author} {\bibfnamefont {H.-L.}\
  \bibnamefont {Kim}}, \bibinfo {author} {\bibfnamefont {H.~C.}\ \bibnamefont
  {Walker}}, \bibinfo {author} {\bibfnamefont {D.}~\bibnamefont {Adroja}},
  \bibinfo {author} {\bibfnamefont {V.}~\bibnamefont {Sechovsk{\'y}}}, \ and\
  \bibinfo {author} {\bibfnamefont {J.-G.}\ \bibnamefont {Park}},\ }\href
  {\doibase 10.1038/s41535-018-0137-9} {\bibfield  {journal} {\bibinfo
  {journal} {npj Quantum Mater.}\ }\textbf {\bibinfo {volume} {3}},\ \bibinfo
  {pages} {63} (\bibinfo {year} {2018})}\BibitemShut {NoStop}%
\bibitem [{\citenamefont {Liu}\ and\ \citenamefont
  {Balents}(2017)}]{PhysRevLett.119.087202}%
  \BibitemOpen
  \bibfield  {author} {\bibinfo {author} {\bibfnamefont {J.}~\bibnamefont
  {Liu}}\ and\ \bibinfo {author} {\bibfnamefont {L.}~\bibnamefont {Balents}},\
  }\href {\doibase 10.1103/PhysRevLett.119.087202} {\bibfield  {journal}
  {\bibinfo  {journal} {Phys. Rev. Lett.}\ }\textbf {\bibinfo {volume} {119}},\
  \bibinfo {pages} {087202} (\bibinfo {year} {2017})}\BibitemShut {NoStop}%
\bibitem [{\citenamefont {Nakatsuji}\ and\ \citenamefont
  {Arita}(2022)}]{doi:10.1146/annurev-conmatphys-031620-103859}%
  \BibitemOpen
  \bibfield  {author} {\bibinfo {author} {\bibfnamefont {S.}~\bibnamefont
  {Nakatsuji}}\ and\ \bibinfo {author} {\bibfnamefont {R.}~\bibnamefont
  {Arita}},\ }\href {\doibase 10.1146/annurev-conmatphys-031620-103859}
  {\bibfield  {journal} {\bibinfo  {journal} {Annual Review of Condensed Matter
  Physics}\ }\textbf {\bibinfo {volume} {13}},\ \bibinfo {pages} {119}
  (\bibinfo {year} {2022})},\ \Eprint
  {http://arxiv.org/abs/https://doi.org/10.1146/annurev-conmatphys-031620-103859}
  {https://doi.org/10.1146/annurev-conmatphys-031620-103859} \BibitemShut
  {NoStop}%
\end{thebibliography}%

\end{document}